\documentclass[printer]{aa}
\usepackage{amssymb,amsmath}
\usepackage{graphicx}
\usepackage{inputenc}
\usepackage{booktabs}
\usepackage{multirow}
\usepackage[caption=false]{subfig}
\usepackage{floatrow}
\titlerunning{Observationally-driven kinetic approach to coronal heating}
\authorrunning{Moraitis et al.}

\begin{document}
\title{An observationally-driven kinetic approach to coronal heating}
\author{K. Moraitis \inst{1} 
\and A. Toutountzi \inst{2}
\and H. Isliker \inst{2}
\and M. Georgoulis \inst{1}
\and L. Vlahos \inst{2}
\and G. Chintzoglou \inst{3}}
\institute{RCAAM, Academy of Athens, 4 Soranou Efesiou Street, 11527 Athens, Greece
\and Department of Physics, Aristotle University, 52124 Thessaloniki, Greece
\and School of Physics, Astronomy, George Mason University, 4400 University Dr., MSN 2F2, Fairfax, VA 22030, USA}

\abstract{}{Coronal heating through the explosive release of magnetic energy remains an open problem in solar physics. Recent hydrodynamical models attempt an investigation by placing swarms of ``nanoflares'' at random sites and times in modeled one-dimensional coronal loops. We investigate the problem in three dimensions, using extrapolated coronal magnetic fields of observed solar active regions.}
{We apply a nonlinear force-free field extrapolation above an observed photospheric magnetogram of NOAA active region (AR) 11158. We then determine the locations, energy contents, and volumes of “unstable” areas, namely areas prone to releasing magnetic energy due to locally accumulated electric current density. Statistical distributions of these volumes and their fractal dimension are inferred, investigating also their dependence on spatial resolution. Further adopting a simple resistivity model, we infer the properties of the fractally distributed electric fields in these volumes. Next, we monitor the evolution of $10^5$ particles (electrons and ions) obeying an initial Maxwellian distribution with a temperature of 10 eV, by following their trajectories and energization when subjected to the resulting electric fields. For computational convenience, the length element of the magnetic-field extrapolation is 1 arcsec, or $\sim$725 km, much coarser than the particles' collisional mean free path in the low corona ($0.1-1$ km).}
{The presence of collisions traps the bulk of the plasma around the unstable volumes, or current sheets (UCS), with only a tail of the distribution gaining substantial energy. Assuming that the distance between UCS is similar to the collisional mean free path we find that the low active-region corona is heated to 100-200 eV, corresponding to temperatures exceeding 2 MK, within tens of seconds for electrons and thousands of seconds for ions.}
{Fractally distributed, nanoflare-triggening fragmented UCS in the active-region corona can heat electrons and ions with minor enhancements of the local resistivity. This statistical result is independent from the nature of the extrapolation and the spatial resolution of the modeled active-region corona. This finding should be coupled with a complete plasma treatment to determine whether a quasi-steady temperature similar to that of the ambient corona can be maintained, either via a kinetic or via a hybrid, kinetic and fluid, plasma treatment. The finding can also be extended to the quiet solar corona, provided that the currently undetected nanoflares are frequent enough to account for the lower (compared to active regions) energy losses in this case.}

\maketitle

\section{Introduction}
The solar corona is a hot (temperature in excess of $10^6$ K), tenuous  ($\approx 10^8-10^9 \rm{cm}^{-3} $), fully ionized plasma. Below the corona, the transition region and the chromosphere are the sites of intrinsically nonlinear dynamical phenomena. The mechanisms heating the coronal plasma and maintaining its temperature above one million K are poorly understood for more than 75 years. The solar coronal heating problem, extending also to Sun-type stars, remains among the main astrophysical puzzles. Although considerable progress has been made, apparent solutions remain highly controversial. A key reason why the problem remains open is that it is ill-posed: while  the nature of coronal heating is assumed to be magnetic, the magnetic field vector is known in only the thin photospheric interface. Hence, the pertinent open questions are: (a) how does the solar atmospheric system (photosphere, chromosphere, transition region and corona) interact and interlink to sustain the observed temperatures? (b) what is the precise role of the magnetic field emerging from the turbulent convection zone and extending into the solar atmosphere to release part of its energy? 

What is already known is that emergence of new magnetic flux and interaction with pre-existing magnetic fields forms complex magnetic topologies which are continuously tangled and shuffled by the photospheric flow field, leading to an inherently complex, nonlinear dissipation of magnetic energy \citep[][and references therein]{Longcop, Archontis}.

The most prominent candidate mechanisms for coronal heating are dissipation of magnetohydrodynamic (MHD) waves and magnetic reconnection \citep[see reviews][]{Klimchuk2006, Klimchuk2015, Pranel2015, Cargill15}, both eligible due to the low value of the $\beta $-parameter in the coronal plasma. However, an attempted  distinction between the two mechanisms is probably artificial, as the dissipation of unstable current sheets (UCS) involves the production of waves and the nonlinear evolution of waves or turbulent eddies in a structured medium can lead to the formation of current sheets \citep{Einaudi96, Georgoulis}. The turbulent convection zone transports energy to the solar atmosphere and into oscillating modes, transient currents and evolving current sheets. Waves and current sheets hence maintain independent roles only in a first-order, linear approximation \citep{Velli}.

Among the leading theories for coronal heating is the well-known Parker conjecture \citep{Parker87, Parker88}, also known as ``nanoflare heating''. The main idea is that strong, localized currents in the solar corona are produced by the braiding of magnetic fields whose photospheric footprints are continuously shuffled. Tangential discontinuities are then formed, where magnetic-field gradients and associated electric currents are steep, signifying the interface between different magnetic-flux systems. These currents may dissipate via magnetic reconnection, by some resistive instability. The released energy is responsible for heating and acceleration of the local ambient plasma, while the post-reconnection magnetic configuration is partly relaxed.  

\citet{Rappazzo07, Rappazzo08} used reduced MHD to investigate the formation and evolution of current sheets and the turbulent cascade of magnetic energy toward small-scale release events  (i.e., ``nanoflares'') in simple magnetic configurations (isolated coronal loops). Their system is continuously driven by the photospheric boundary and leads to the dissipation of  numerous small-scale current sheets \citep[see also][]{Rappazzo10, Velli}.

It is clear that a key aspect of coronal heating is hidden in the unknown magnetic-field properties above the photosphere and its evolution due to the turbulent photospheric interface. In a series of articles, the formation of null points, separators and separatrix surfaces are discussed and associated with coronal heating and flaring in active regions \citep[see the review of ][]{Parnell2}. Most of the analyses presented so far relied on relatively simple magnetic configurations, initially based on independent or interacting magnetic loops (magnetic threads). Nonetheless, the dynamical evolution of even simple structures soon leads to extremely complex magnetic fields, where small scales dominate both the evolution of the system and the energy-dissipation process \citep{Bowness, Tam}.

An interesting development is the use of three-dimensional (3D) MHD simulations for the understanding of both nanoflares and coronal heating \citep{Peter04, Gudiksen, Binger, Hansteen1}. Several authors have discussed the limits of their MHD simulations based on simple arguments \citep[see also the review by ][]{Peter1}. For a coronal temperature of  $10^6$ K and a coronal density of $10^9 \rm{cm}^{-3}$  the mean free path of plasma particles is 0.1 - 1 km. At such small scales, special care should be taken to ensure validity of the MHD approximation. It is also well known that UCS have dimensions of the order of the mean free path. Hence, their evolution can be followed by 3D particle-in-cell simulations. The use of adaptive grids in large-scale simulations of the solar corona is difficult because the fragmentation of electric currents soon drops below even the finest spatial resolution. It is then obvious that a kinetic approach combined with the MHD methodology is necessary to capture the coronal heating process. Another fundamental question is then borne, namely ``how do large scales evolve when the main dissipation processes (reconnection and waves) are operating on sub-resolution scales?'' \citep{Cargill13}.

A simple method to detect MHD discontinuities and magnetic reconnection in the solar wind was employed \citep{Greco1, Greco2, Servidio, Osman}. They used the Partial Variance of Increments method \citep{Greco1} applied to either MHD simulation results or in-situ magnetic-field measurements. The reconnection events and current sheets were found to be concentrated in intervals of intermittent turbulence.

\citet{Vlahos04} followed a different approach, aiming to identify unstable magnetic discontinuities and reconnection in complex active-region magnetic configurations. They based their analysis on linear force-free extrapolated magnetic fields of the active-region corona, aiming to detect discontinuities and subsequent electric-current fragmentation. A key finding was that flaring and non-flaring active regions share similar statistical distributions of unstable volumes, with available energies in agreement with the observed occurrence frequency distributions of solar flares, (i.e., robust power laws with similar indices, of the order $\sim 1.5$).

We follow a similar reasoning in this work, although by adopting a different approach: we investigate the statistical properties of the fragmented distributions of tangential discontinuities and UCS. The magnetic field of the active-region corona is reconstructed using a nonlinear force-free extrapolation method, validated by recently published techniques. The target is the repeatedly eruptive NOAA AR 11158, studied in literally dozens of works, none of which in this light, however. Section 2 describes the extrapolation method and the statistical properties of UCS, comparing also the results with those of previous studies. In Section 3 we release a large number of electrons and ions in random locations, with energies following a Maxwellian distribution, and follow their heating and acceleration in connection with the statistical properties of the extrapolated fields. In Section 4 we summarize the key points of our analysis and present our main results and conclusions.

\section{Non-linear magnetic field extrapolation and the statistics of current fragmentation}

We focused on a well-studied active region (AR), NOAA AR 11158 \citep[e.g.][]{schrij11,jiang12,sun12,liu12,vemar12,chintzo13}. This AR gave the first X-class flare of the current solar cycle (X2.2 on February 15, 2011  01:44 UT) as well as many M- and C-class flares during the interval February 11-16, 2011. The AR evolved from a simple, bipolar, to a complex, quadrupolar, structure with an enhanced and strongly sheared magnetic polarity inversion line (PIL). The total unsigned magnetic flux of the AR was increasing during this period and reached a maximum value of $\sim 6\times 10^{22}$\,Mx \citep{tgl13}.

The input for our analysis were photospheric vector magnetogram cutouts from the Helioseismic and Magnetic Imager (HMI; \citealt{scherrer12}) onboard the \textit{Solar Dynamics Observatory}. HMI magnetograms were already transformed into cylindrical equal area projections which preserve the pixel area and treated for the azimuthal 180$^{\rm o}$ ambiguity as described in \citet{Hoek}. The pixel size is $\sim$0.36\,Mm ($\sim 0".5$) and the size of the cutouts 600$\times$600 pixels. We selected two snapshots from the AR evolution, one on February 13, 2011 03:58 UT and another one on February 14, 2011 21:58 UT, for reasons that will be explained in the following section.

\subsection{Magnetic Field Extrapolation}

A force-free magnetic field model of the solar corona dictates field-aligned electric currents by a valid (i.e., divergence-free) magnetic field solution. Hence,

\begin{equation} \label{FF}
\nabla\times\mathbf{B} =\alpha\,\mathbf{B}\ ; \ \  \nabla\cdot\mathbf{B} =0
\end{equation}
\\
where $\alpha$ is the force-free parameter, generally varying as a function of position but remaining constant along a given magnetic field line. This is the case of Non-Linear Force-Free (NLFF) fields, which has generally advanced our understanding of the overall morphology of ARs and current channels along PILs \citep[][and references therein]{wiegsakurai}. A special case is the linear force-free field, in which $\alpha$ is constant throughout the extrapolated volume, but when the extrapolated magnetic field lines are compared to structures from EUV images \citep{wiegsakurai}, the linear solution is often found to fail in recovering the overall magnetic field topology.
For our investigations we used an optimization technique \citep{wieg04} for computing the 3D NLFF field in the corona. Using appropriate boundary conditions, this numerical method yields a NLFF field solution by minimizing a penalty function, $L$, in the computational volume, $V$, as
\begin{equation}
L= \int \limits_V \, w(x,y,z)[B^{-2}|(\nabla\times \mathbf{B})\times \mathbf{B}|^{2}+|\nabla \cdot \mathbf{B}|^{2}]\, \mathrm{d} V
\end{equation}\\
where $w(x,y,z)$ is a scalar function with a value of 1 in the physical domain of the volume that drops smoothly to zero when approaching the top and lateral boundaries. When $L=0$, both the Lorentz force is zero and the solenoidal condition is satisfied in the entire computational volume, which then contains a perfect NLFF field for a force-free photospheric boundary. As we will see below, numerical imperfections are inevitable in practice.

\begin{figure*}[ht]
\centering
\includegraphics[width=17cm]{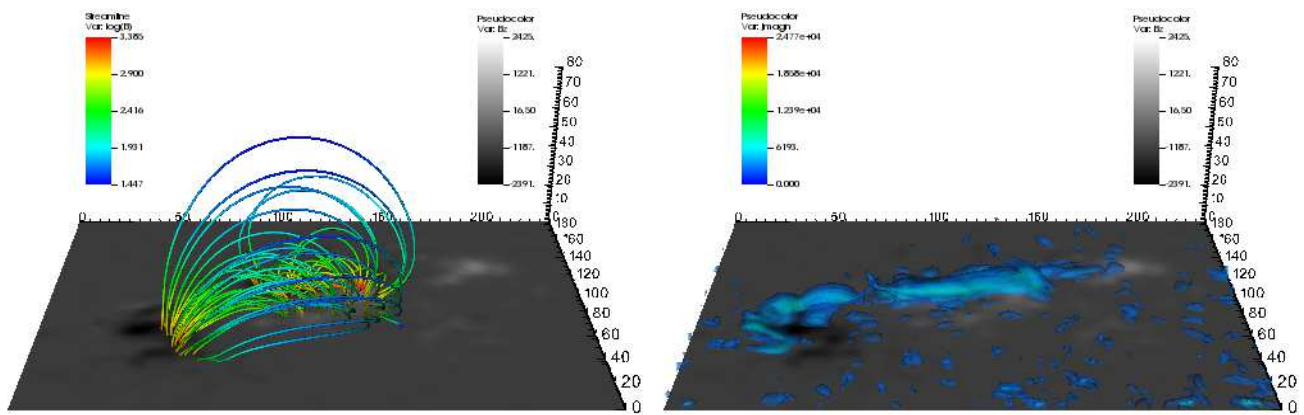}
\caption{Original magnetogram for the eruptive NOAA AR 11158 on February 14, 2011 at 21:58 UT together with the NLFF field lines (left), and zoomed view with isosurfaces of the current density (right) showing the low-lying current fragmentation.}
\label{figdata}
\end{figure*}

We chose to create models of the 3D coronal field in two resolutions for each time-frame. Starting from the original 600$\times$600 images we trimmed quiet-Sun regions to reduce the size to 500$\times$400 pixels, still containing the entire AR. We then rebinned these by factors of 2 and 4 to reach grid sizes equal to 250$\times$200$\times$128 and 125$\times$100$\times$64 pixels, respectively. In the following we will refer to these resolutions as `1"' and `2"' respectively. In all cases the vector data were pre-processed according to \citet{wieg06} so that they are consistent with the assumption of a force-free magnetic field. An example of the resulting 3D NLFF field is shown in Fig.~\ref{figdata}.


\subsection{Statistics of current fragmentation}
The overall AR magnetic field configuration shown in Fig.~\ref{figdata} was used to further study the coronal volumes above the ARs, in terms of their energetic content and certain statistical properties (pertinent to our particle model). As discussed in the Introduction, this structure is comprised of current-carrying unstable volumes, formed as an excess of magnetic energy accumulates near regions where magnetic gradients build-up, just before its release via reconnection. The identification of these regions was based on the currents associated with the tangential magnetic discontinuities built in the reconstructed NLFF field.

The free energy, which powers solar dissipative events, is the excess energy of a magnetic field $\mathbf{B}$ that occupies a volume $\mathcal{V}$ relative to that of the current-free (known as potential) magnetic field $\mathbf{B}_\mathrm{p}$ in the same volume. Free energy can be given by either of the expressions
\begin{eqnarray}\label{ecdef}
E_\mathrm{f1}&=&\frac{1}{8\pi}\int_\mathcal{V} \mathrm{d}V\,\mathbf{B}^2-\frac{1}{8\pi}\int_\mathcal{V} \mathrm{d}V\,\mathbf{B}_\mathrm{p}^2 \label{ecdef1}\\
E_\mathrm{f2}&=&\frac{1}{8\pi}\int_\mathcal{V} \mathrm{d}V\,(\mathbf{B}-\mathbf{B}_\mathrm{p})^2,\label{ecdef2}
\end{eqnarray}
which are equivalent for exactly solenoidal field, and for potential field that has the same normal components with the given one on the volume's boundaries \citep{moraitis14}. Any discrepancy between Eqs.~(\ref{ecdef}) and (\ref{ecdef2}) thus provides means to perform a validation check on the extrapolated magnetic field or, in other words, how solenoidal the three-dimensional field solution actualy is. To quantify this, we define the free-energy accuracy factor $r$ similarly to a quantity used in \citet{su14} for the same purpose, namely as
\begin{equation}
r=\frac{\left| E_{\rm f1}-E_{\rm f2}\right| }{\left| E_{\rm f1}\right|+\left| E_{\rm f2} \right|}.
\label{eq:r}
\end{equation}
This quantity provides a dimensionless relative error in free energy due to non-solenoidality in the field. By construction, it is $0\leq r\leq 1$ with $r=0$ denoting perfect solenoidality (except in the unlikely case where $\int \mathrm{d}V\,\mathbf{B}_\mathrm{p} \cdot (\mathbf{B}-\mathbf{B}_\mathrm{p})=0$, for $\mathbf{B}\neq \mathbf{B}_\mathrm{p}$), while $r=1$ denotes an unphysical, negative $E_{f1}$. Apart from an average value for $r$ in the extrapolated volume, we also define the local value for this accuracy factor at each cubic voxel which we denote as $r_i$. This local factor $r_i$ has no physical meaning since its nominator depends on the non-local divergence of the field. However, it can be used to discriminate regions where the extrapolated field is of low-quality, as we show in the following.

The global value of this free-energy accuracy factor for the whole extrapolated field was also the reason for choosing the specific snapshots of AR 11158's evolution from the set of NLFF fields that were used in \citet{moraitis14}. The first snapshot (February 13, 2011 03:58 UT) corresponds to the absolute minimum of $r$ during the evolution of the AR with a value of $r=0.13$, while the second (February 14, 2011 21:58 UT) ranks third with $r=0.22$. Although these values may not seem low, the quality of the fields at these instances is adequately high as other, more common metrics indicate. For example, the average absolute fractional flux increase \citep{wheatl00} for the high-resolution Feb 13 and Feb 14 snapshots are $\left\langle \left| f_i \right| \right\rangle =8.6\,10^{-4}$ and $\left\langle \left| f_i \right| \right\rangle =7.6\,10^{-4}$, respectively. Additionally, the force-free metric of the average Lorentz force relative to its components \citep{malanus14} is quite low, $\xi\simeq 0.06$, for both snapshots. 

Albeit not precisely solenoidal, our magnetic field solution shows solenoidality that is comparable to the one of extrapolations used in previous studies \citep[e.g.][]{derosa15}. We therefore use these extrapolation results, adding that we are interested particularly in the UCS distribution and not in the field-line connectivity, that would be impacted by imperfections in the divergence-free condition. A more thorough extrapolation investigation that might produce more accurate solenoidal and force-free fields exceeds the scope of this work.

In anticipation of the X-class flare on February 15, 2011 01:44 UT, both snapshots refer to the preflare phase in AR 11158. The first one follows a 20h period of fast flux emergence and the formation of a filament along the main active-region PIL. During the 30 hours elapsing between the first and the second snapshot, magnetic flux keeps emerging, albeit at a decreased rate, but the current density build-up in the lower layers of the coronal volume  continues, as the filament is stretched and smaller-scale eruptions take place \citep{sun12}. Both snapshots were therefore also selected so as to not coincide with any of the major flares/CMEs, with the major X-class flare occurring 3 hours after the second snapshot.

Before proceeding with the current fragmentation analysis, we perform a simple check on the electrical currents present in the volume. Starting from Amp\'ere's law
\begin{equation}\label{ampere}
\mathbf{j}=\frac{c}{4\pi} \nabla \times \mathbf{B},
\end{equation}
we calculate the electric current density $\mathbf{j}$, using the appropriate (forward, backward or centered) second-order difference operators for the field's rotation. In Fig.~\ref{figdata} we show the 3D structure of the currents and notice that the AR corona is nearly filled by accumulated electric currents that (1) stay relatively close to the photospheric boundary (up to $\sim 10$~Mm) and (2) are mostly concentrated along the AR's PIL.

\begin{figure}[ht]
\centering
\subfloat[][]{\includegraphics[width=0.45\columnwidth]{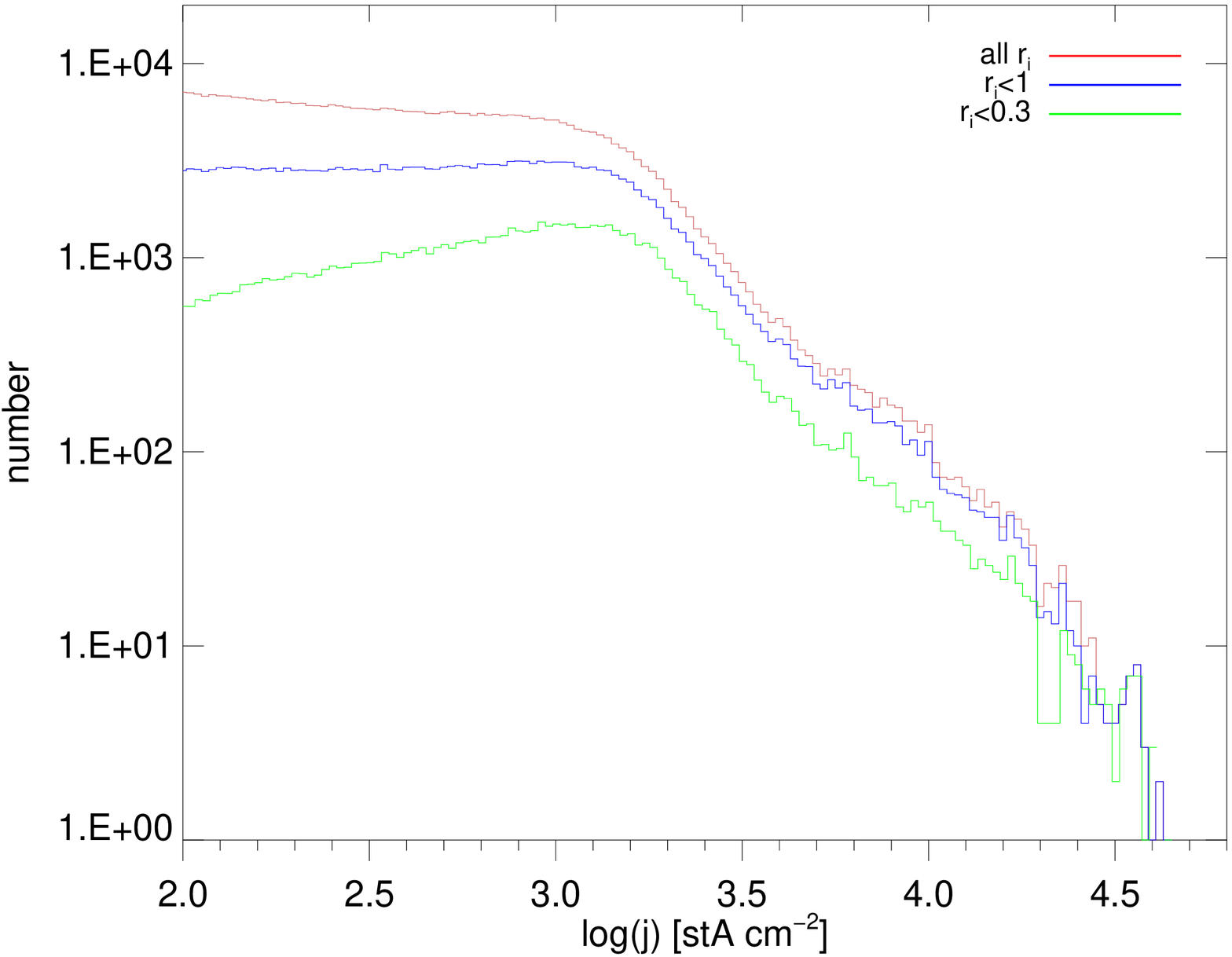}%
		\label{fig:2a}}
\subfloat[][]{\includegraphics[width=0.46\columnwidth]{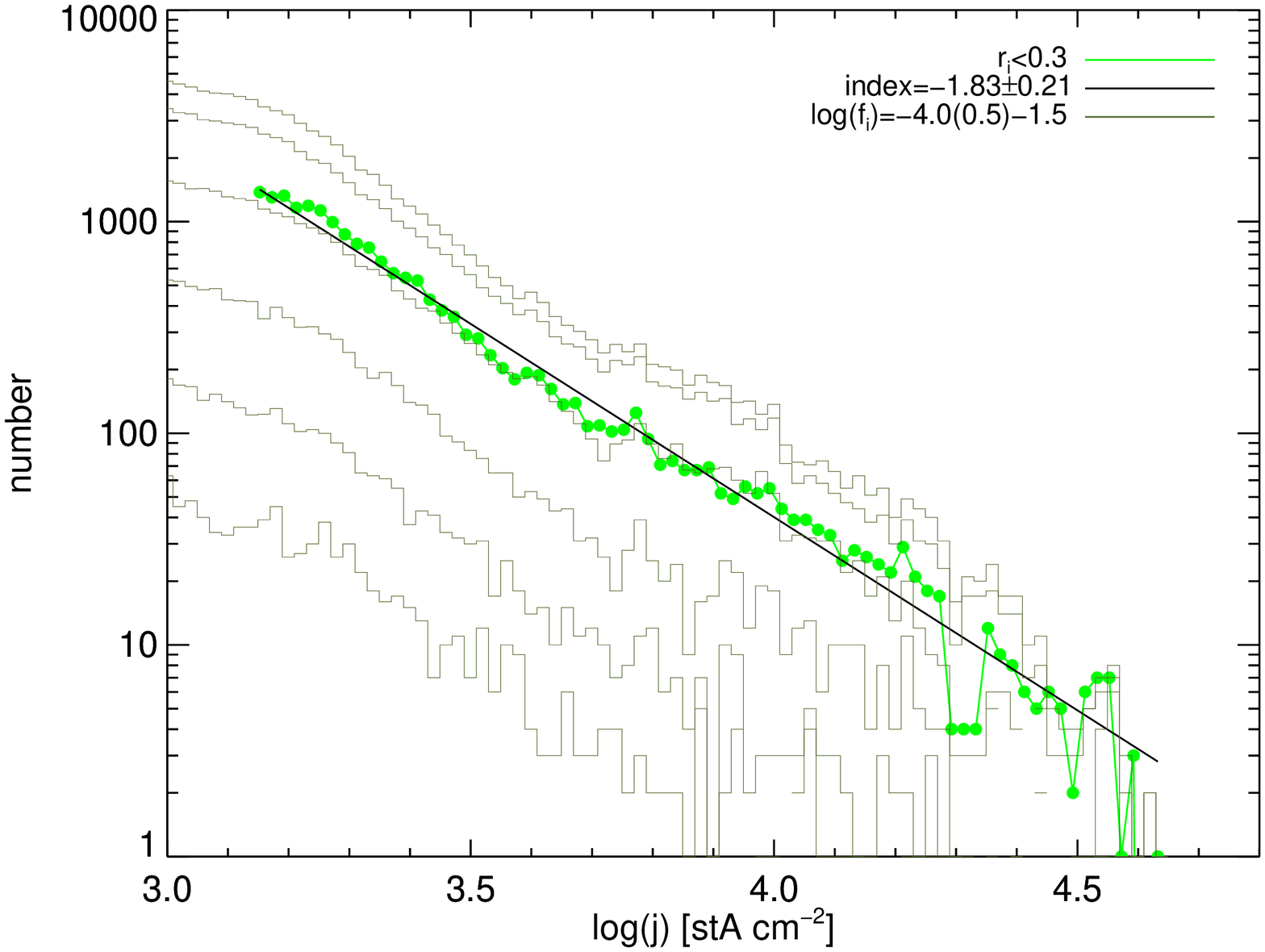}%
		\label{fig:2b}}
\caption{Logarithmic histogram of the electric current density for different threshold values of the free-energy accuracy factor \protect\subref{fig:2a}, and zoom of the $r_i<0.3$ curve and power law fit to it \protect\subref{fig:2b} for the snapshot of February 13, 2011 at 03:58 UT. Also shown are histograms of the current density for different values of the absolute fractional flux increase, $\left| f_i \right|$.}\label{figt2}
\end{figure}

We then construct a histogram of the magnitude of the current, $j$, that is shown in Fig.~\ref{figt2}. The higher-current part of the distribution exhibits a power-law that does not change even if we exclude points where the 3D extrapolated field is of poor quality (as indicated by the free-energy factor $r_i$). In the following we consider only the highest-quality points with $r_i<0.3$. This is further justified by noting that the $r_i<0.3$ curve in Fig.~\ref{figt2}b is very close to the one for $\left| f_i \right|<10^{-2.5}$.

The next step is to determine the threshold, $j_\mathrm{th}$, above which the distribution is a power law and the corresponding index $q$. This is done by fitting a finite power-law with a lower-values departure to the current distribution and identifying the location of the power-law breakdown with the threshold and the power-law index above it with $q$. The results of the fitting procedure, and also the maximum value of the current $j_\mathrm{max}$, are reported in Table~\ref{tab1} for the two snapshots and the two grid sizes considered.

For the distribution of the high-resolution snapshot of February 13, 2011 at 03:58 UT that is shown in Fig.~\ref{figt2} we estimate a power-law index of $1.83\pm 0.21$ above the threshold $j_\mathrm{th}=1.38 \times 10^3\,\mathrm{stA}\,\mathrm{cm}^{-2}$. A general trend seen in Table~\ref{tab1} is that the (absolute) power-law index is an increasing function of resolution, a result valid for both snapshots. The same is true for the threshold and maximum values of the current, their ratio however, i.e. the span of the power law, is less affected by the spatial resolution, since $j_\mathrm{th}/j_\mathrm{max}=(0.030-0.065)$ in all cases. Another finding is that a power law fit is more suitable for the first snapshot than for the second (judging from the values of $\chi^2$ and the form of the distributions), and also for each high-resolution snapshot than for the low-resolution ones, and thus a different function could describe these distributions better. However we limit our analysis to the general, single power-law case.

In the remainder of this work we use only the highest-resolution data of the snapshots of February 13, 2011 at 03:58 UT and February 14, 2011 at 21:58 UT, and comment on the effect of resolution on our results in Section 4.

\begin{table*}[ht]
\centering
\caption{Power-law fitting parameters for the high-current tail.}
\begin{tabular}{cccccc}
\toprule
snapshot & pixel size (") & $q$ & $j_\mathrm{th}\,(\mathrm{stA}\,\mathrm{cm}^{-2})$ & $j_\mathrm{max}\,(\mathrm{stA}\,\mathrm{cm}^{-2})$ & $\chi^2$ (dof\tablefootmark{*}) \\
\midrule
\multirow{2}{3em}{Feb 13} & 2 & $1.37\pm 0.18$ & 4.50E2 & 6.97E3 & 2.04 (22) \\
& 1 & $1.83\pm 0.21$ & 1.38E3 & 4.34E4 & 0.45 (28) \\
\midrule
\multirow{2}{3em}{Feb 14} & 2 & $0.76\pm 0.13$ & 3.20E2 & 1.22E4 & 3.36 (30) \\
& 1 & $1.59\pm 0.14$ & 1.58E3 & 2.48E4 & 0.93 (20) \\
\bottomrule
\end{tabular}
\tablefoottext{*}{degrees of freedom}
\label{tab1}
\end{table*}

In order to locate the adjacent regions forming spatially disjoint groups (clusters) which we associate with UCS, we implement a partitional, hard clustering algorithm that uses the Manhattan distance as a dissimilarity measure \citep{gan11} to group 3D voxels into clusters. Under the Manhattan distance, a cluster is defined as a group of points, each connected to at least one other group member, called its immediate neighbor, via a single edge of the rectangular lattice underlying the 3D volume or, as a limiting case, a single point isolated in the sense just described.

\begin{figure*}[ht]
\sidecaption
\centering
\includegraphics[width=12cm,clip]{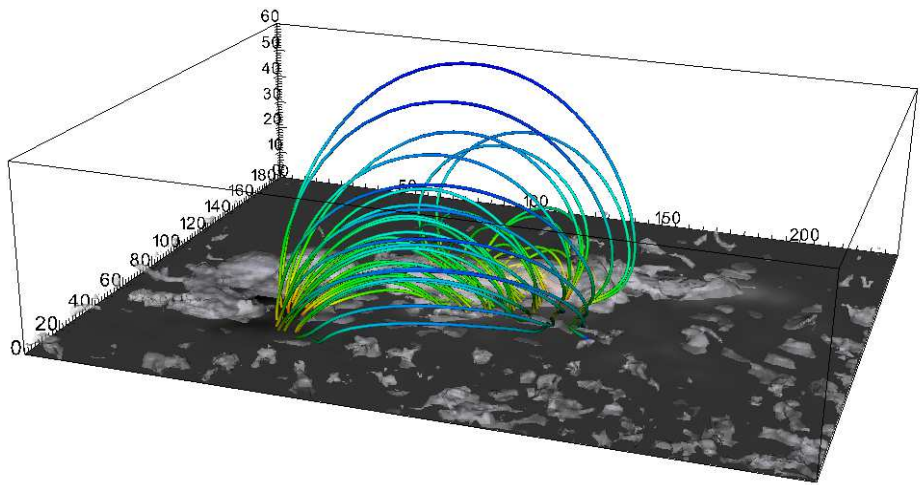} 
\caption{The clustering scheme recovered for the February 14 snapshot.}\label{cartoonish}
\end{figure*}

The group of 3D pixels on which clustering was applied exhibits current densities exceeding a threshold $j_{\rm th}$, determined in Table~\ref{tab1}. It is worthwhile noting that the number of clusters identified was near $\sim7700$ for the February 13 snapshot and $\sim8600$ for the February 14 one, all gathered within lower altitudes in the AR (below $z\sim 12 - 13$ pixels, up to $\sim 9\, \mathrm{Mm}$ above the photosphere, for both frames, Fig. \ref{figdata}). The clustering scheme recovered for the February 14 snapshot is shown in Figure \ref{cartoonish}, where the cluster sizes range between 1 and 600 voxels. The fragmentation observed in most of the areas contrasts a couple of large clusters, warning about a possible bias introduced in the form of these outliers, to the statistical analysis following. Nevertheless, we include all clusters in the subsequent analysis.

From this point on, all subsequent analysis is performed on a cluster-based level. It is, therefore, convenient to define a point representing each cluster. This point was chosen to be the centroid of a cluster, defined here as the point located at the weighted (by the current density magnitude per pixel) average position of all the points comprising the cluster. Specifically, if a cluster consists of $N_C$ points with position vectors $\mathbf{P}_m,\;m=1,\ldots, N_C$, then the position vector of its centroid $\mathbf{P}_C$ is defined as
\begin{equation}\label{centroid}
\vec{P}_C = \frac{\sum_{m=1}^{N_C} j_m  \vec{P}_m}{\sum_{m=1}^{N_C} j_m},
\end{equation}
where $j_m$ is the current density magnitude of an individual voxel.

We refer to clusters of contiguous voxels as UCS. In the following, we will study PDFs of the UCS free energy, volume, and average current density.

\begin{figure}[ht]
\centering
\includegraphics[width=0.95\textwidth]{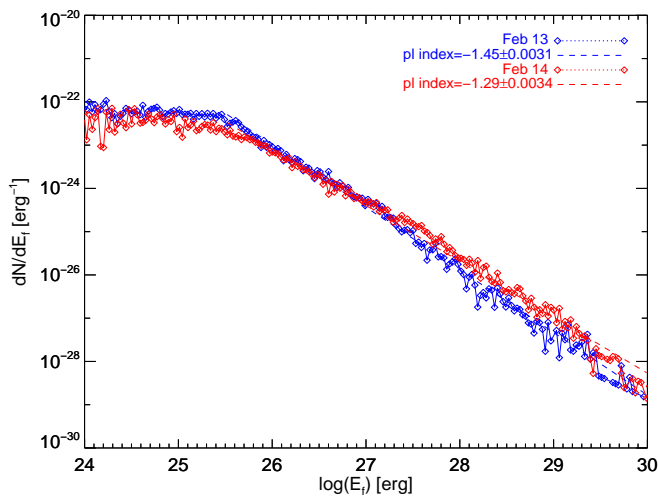}
\caption{Histograms of the free magnetic energy of UCS based on the clustering identification process performed on both high-resolution snapshots. Notice the decrease in the absolute power law index from the February 13 (blue), to the February 14 (red) snapshot. The corresponding fits use the respective colors while legends provide the power-law (pl) index for each snapshot.}\label{energy}
\end{figure}

\paragraph{Energy distribution.$-$}\label{FractalPAR}
As a second step of the analysis on the current fragmentation exhibited by NOAA AR 11158, the free magnetic energy per UCS is calculated as the sum of the free magnetic energy of the points comprising the cluster and its distribution is computed using the approach known as logarithmic binning \citep[see e.g.][]{Newman05,Clauset09}. In Fig. \ref{energy} the results for both snapshots are presented, overlaid by their respective power law fits seen to span at least four orders of magnitude before their scaling breakes down below $\sim 5\times 10^{24}\,\mathrm{erg}$, i.e. while approaching the known limit imposed by resolution constraints. The absolute power law index flattens from 1.45 on February 13 to 1.29 on the February 14 snapshot, in accordance to the observation already made for the 1" cutouts regarding the current distribution of the individual 3D pixels (see Table \ref{tab1}). The same behavior will be recovered later for the other quantities.

\begin{figure}[ht]
\centering
\includegraphics[width=0.95\textwidth]{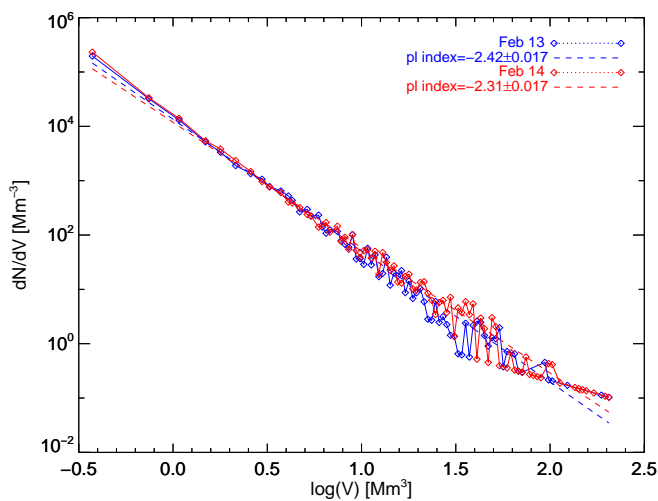}
\caption{Distributions of UCS volumes for the two snapshots on February 13 and 14, 2011, using the same color notation with  Fig. \ref{energy}. As all observations related to the high-resolution snapshots, the power-law fitted distribution of the volumes flattens from February 13 to February 14.} \label{volume}
\end{figure}

\paragraph{Volumes Distribution.$-$}\label{VolumesPAR}
Extracting information regarding the spatial extent of the dissipation regions regions after clustering is also a straightforward process. One can simply ``log''-bin the number of points $N_C$ comprising each of the clusters recovered into a logarithmic histogram or, equivalently, denote the volume of any of these clusters as $V$ and bin these volumes instead. Adopting the latter approach, the distribution functions of values for the two high-resolution snapshots are shown in Fig. \ref{volume}. Respective power-law fits are superposed.

\paragraph{Fractal Dimension.$-$}\label{FractalPAR2}
As a further basic characteristic of the cluster distribution, we determine its fractal dimension,
which will be also needed in the study of the particle dynamics. We employ the method based on the correlation sum using the algorithm introduced by \citet{GrassProcc1983a, GrassProcc1983b}. The correlation sum $\widehat{C}(r)$ is given by
\begin{equation}\label{gen_corr_sum}
\widehat{C}(r) =   \frac{1}{N} \sum_{i=1}^{N}   \frac{n_i(r)}{N-1}   ,
\end{equation}
where 
\begin{equation}\label{inner_corr_sum}
n_i(r) = \sum_{j = 1,\,j \neq i}^{N} \Theta \left(  r - | \mathbf{P}^i - \mathbf{P}^j |  \right).
\end{equation}
In the equations above, $n_i(r)$ is the number of points $\mathbf{P}^j$ other than $\mathbf{P}^i$, contained in a sphere of radius $r$ centered at point $\mathbf{P}^i,\, i=1,2,\ldots,N$, where $N$ is the total number of points, and $\Theta$ is the Heaviside function with  $|\cdot |$ denoting the Euclidean distance between points $\mathbf{P}^i$ and $\mathbf{P}^j$. In this particular application, by point we mean the centroid (Eq. \ref{centroid}) of the cluster identified as an unstable volume, and all distances are measured in pixel units.

Given the above relations (Eqs. \ref{gen_corr_sum} and \ref{inner_corr_sum}), a measure of the structure of a fractal known as the correlation dimension $D_F$ \citep[see also][]{Grass1983} is defined as
\begin{equation}\label{corr_dim}
D_F = \lim_{r \rightarrow 0} \frac{\ln \widehat{C}(r)}{\ln r}.
\end{equation}
Quantifying the complexity of the structures under consideration with this particular definition of the fractal dimension allows for fast and more efficient computations (as compared to e.g. the box-counting method, \citet{GrassProcc1983a}).

\begin{figure}[ht]
\centering
\includegraphics[width=0.95\textwidth]{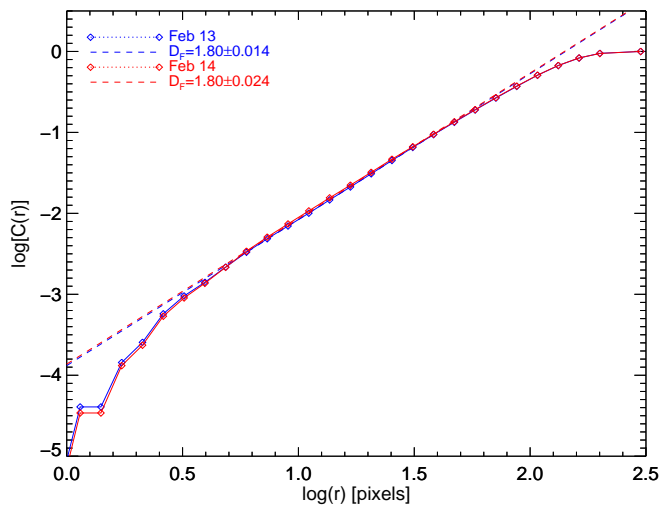}
\caption{The correlation dimension $D_F$ as estimated by a linear fit to the correlation sum $\widehat{C}(r)$ vs. $r$. For the two snapshots on February 13 and 14, 2011 the fractal dimension of the structure formed by the magnetic discontinuities identified and systematically recorded, is found to remain constant in time, around the value of $1.8$.}\label{fractal}
\end{figure}

In Fig. \ref{fractal}, the logarithmic plot of the correlation sum $\widehat{C}(r)$ is shown as a function of the radius of the sphere $r$. Employing the original and most straightforward approach to compute the correlation dimension, a linear fit to the plots in Fig. \ref{fractal} gives $D_F$ according to equation (\ref{corr_dim}), yielding values close to 1.8 in both cases of data sets. Even though the power law scaling (bounded from above due to the finite size of
the sets) breaks down at around the minimum inter-point distance, the correlation dimension
suitably probes the intermediate-to-small $r$-regime.

\begin{figure}[!ht]
\centering
\includegraphics[width=0.95\textwidth]{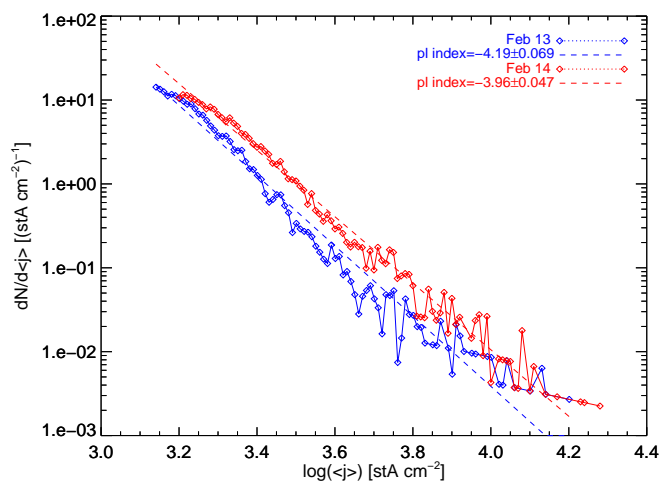}
\caption{Logarithmic histogram of the average current density per cluster, for the high-resolution frames on February 13 and 14, 2011.}\label{current}
\end{figure}

\paragraph{Average Current Density Distribution.$-$}\label{CurrentsPAR}
The energy content of the UCS can be examined by either the free magnetic energy $E_{\rm f}$ per UCS already seen previously or, as it will actually be utilized in the next section, the average current density magnitude $\langle j \rangle$ per UCS. The logarithmic histograms of $\langle j \rangle$ are shown for both snapshots in Fig. \ref{current}. 
We again notice that the power-law distribution flattens from the first to the second frame, as all power laws presented in Figs. \ref{energy} and \ref{volume}. The results of the analysis for all quantities and for both snapshots are shown in Table \ref{tab2}.

\begin{table*}[ht]
\centering
\caption{Synoptic results of the statistical analysis performed on the clustered data, for the two snapshots on February 13 and 14.}
\begin{tabular}{ccc|cc}
\toprule
snapshot & Feb 13 & Feb 14 & Feb 13 -- 150 & Feb 13 -- \textit{Hinode} \\
\midrule
$E_{\mathrm{f}}$ & $1.45\pm 0.003$ & $1.29\pm 0.003$ & $1.27\pm 0.012$ & $1.54\pm 0.004$ \\
$V$ & $2.42\pm 0.02$ & $2.31\pm 0.02$ & $2.32\pm 0.02$ & $2.18\pm 0.02$ \\
$D_F$ & $1.80\pm 0.01$ & $1.80\pm 0.02$ & $1.80\pm 0.03$ & $1.83\pm 0.07$ \\
$\langle j \rangle$ & $4.19\pm 0.07$ & $3.96\pm 0.05$ & $3.65\pm 0.07$ & $3.63\pm 0.05$ \\
\bottomrule
\end{tabular}
\tablefoot{
With the exception of the fractal dimension $D_F$, the rest of the numbers given are the absolute power law indices (with their standard errors) of the performed fits. A reference column with very high-resolution data from the \textit{Hinode} mission for the first snapshot is also provided.
}
\label{tab2}
\end{table*}

\begin{figure}[ht]
\centering
\includegraphics[width=0.95\textwidth]{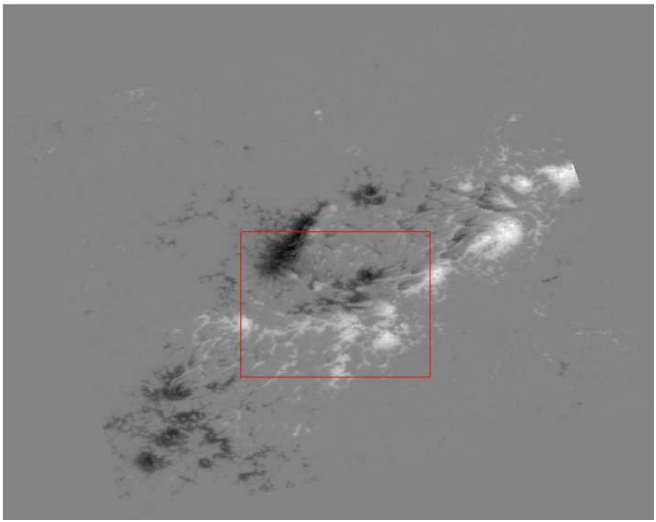}
\caption{\textit{Hinode} magnetogram of February 13, 2011 at 04:00 UT, after transforming the original 512$\times$512 pixels image (with center at central meridian distance and latitude of -12.96$^o$ and -19.94$^o$ respectively) into heliographic coordinates, and area selected for analysis (red box).}\label{hinode}
\end{figure}

To adress the effect of resolution on these results we examine, in addition to the 1" and 2" resolutions of HMI data, Solar Optical Telescope/Spectro Polarimeter (SOT/SP) data from the \textit{Hinode} mission \citep{lites08}, exhibiting a resolution of $\sim$~0".3, for the February 13 snapshot. 
Since the analysis of the entire AR under such a resolution is beyond our computational resources, we restrict it to the 180$\times$180 pixels area depicted by a red box in Fig.~\ref{hinode}. Indeed, a factor-of-three difference between the HMI and the SOT/SP magnetographs would imply a factor of $\sim 3^3$ additional computing nodes in the three-dimensional extrapolation volume. Given that optimization methods in NLFF field extrapolation scale by $\sim N^5$, where $N$ is the number of computing nodes \citep[e.g.][]{schrij06}, an increase in resolution by a factor of 3 would imply an increase in computations by a factor of $\sim 3^{15}\simeq 1.4\,10^7$. We have not performed a rigorous testing of this effect but we maintain that it would be impractical to use the SOT/SP resolution for the entire AR with our resources.

The box in Fig.~\ref{hinode} is chosen so as to be flux-balanced for the extrapolation and to contain the main part of the AR’s PIL. Obviously, the box does not contain the entire AR, so the magnetic-field connectivity will not be identical to the one inferred by extrapolating the entire AR. This said, we are interested in the statistical properties of the extrapolated field, namely the UCS spatial distribution – not their actual locations – that should be fairly insensitive to the spatial resolution for a fractal system such as the one studied here. Moreover, we focus on the AR’s PIL because it is in this area that we expect the most significant UCS to be present. Upon selecting the box, we perform a NLFF field extrapolation of the magnetic field up to a height of $\sim$~27~Mm above the photosphere. After verifying that the extrapolated field is sufficiently divergence- and force-free we follow the same analysis. That is, we calculate the electric current, select only the highest-quality points that show $r_i<0.3$, and then determine the clusters formed by these points and their statistical properties. We find that the fractal dimension is practically unaffected by resolution and remains around 1.8, within uncertainties. The volumes, average current, and free energy distributions show small variations in their power-law indices, in the ranges 2.2-2.4, 3.6-4.2, and 1.3-1.5, respectively.

Since the two snapshots exhibit similar statistical characteristics, we choose to continue treating particle dynamics in the next section using the distributions estimated for the second snapshot, on February 14, 2011.

\section{Particle dynamics in a fractal distribution of fragmented currents}

The motion of charged particles inside an environment of randomly distributed UCS can be analyzed with the use of the two PDFs, namely $P(V)$ and $P(\left\langle j\right\rangle)$, determined in the previous section and the fractal dimension $D_F$, following methods developed by \cite{Vla05}. The charged particle (electron or ion) starts at a random point inside the AR with a random velocity $u_i$ along the magnetic field lines. The initial velocity distribution of the particles is a Maxwellian with initial temperature $T \approx 10^5\,$K (10 eV). The ambient density of the  particles in the low corona is approximately $10^{9}\,$cm$^{-3}$. The charged particle moves freely along a distance $s$ estimated from the fractal dimension $D_F$  (see details below) until it reaches a current sheet where it is energized  by the electric field. This is estimated by Ohm's law $E=\eta \left\langle j\right\rangle$, where $\eta$ is the local resistivity and $\left\langle j\right\rangle$ is estimated from the probability distribution $P(\left\langle j\right\rangle)$ shown in Fig. \ref{current}. Since the free travel of particles is longer than the collisional mean free path we include collisional losses in our analysis. We follow the evolution of the initial particle distribution  in successive time intervals. Let us discuss briefly below the way we reconstruct the dynamic evolution of a distribution of particles inside a fractal distribution of UCS.

\paragraph{Free travel distance.$-$}\label{ftt}
As it was pointed out by \cite{Isliker03}, the probability of a particle, starting at an UCS in the AR, to travel freely a distance $s$ before meeting again an UCS is
\begin{equation}\label{frac}
P(s)=\frac{D_F-2}{s_{\rm max}^{D_F-2}-s_{\rm min}^{D_F-2}}s^{D_F-3}\;\;\;\; s_{\rm min}<s<s_{\rm max}
\end{equation}
if the UCSs are fractally distributed {(with $s_{\rm min}$ and $s_{\rm max}$ the minimum and maximum free travel distance, respectively). This formula is an approximation that applies if $ D_F$ is strictly smaller than 2, as for the case examined here (the corresponding expressions for the cases $D_F=2$ and $D_F>2$ are different and not cited here). From Fig. \ref{fractal}, we estimate the fractal dimension as $D_F=1.8$ (see Table \ref{tab2}), and we assume that $s_{\rm min}=10^8\,$cm, which is close to the resolution of the extrapolations' grid, and $s_{\rm max}=10^{10}\,$cm. Using the distribution $P(s)$, we generate sequences of random free travel distances $s_i$.

\paragraph{Collisional losses.$-$}\label{parcollisions}
The electron and ion Coulomb collision frequency 
is given by
\begin{equation}\label{colfr}
\nu_e = \frac{4\pi n e^4 \ln\Lambda }{m_j^2 U_j^3}
\end{equation}
where $e$ is the elementary charge, $j={e,i}$ corresponds to electrons (e) and ions (i), $m_j$ is the particle mass, $\Lambda$ is the Coulomb logarithm  \citep[see e.g.][]{Karney86}, $n$ is the number density and $U_j$ is the thermal velocity. For the typical values of number density $n=10^{9}\,{\rm cm}^{-3}$ and a temperature of  $T_e=10$~eV, the thermal velocity is $U_e\approx 10^8\,\mathrm{cm}\,\mathrm{s}^{-1}$, so the mean free path is $\lambda_{\rm mfp}\approx 10^4 \,$cm $- 10^5\,$cm. The particles thus lose or gain energy as they travel between current sheets located at distances $10^8\,{\rm cm} <s_i<10^{10}\,$cm that are much larger than the mean free path. 

\paragraph{Acceleration length.$-$}\label{al}
Assuming that the particles interact with a current sheet with volume $V_j=\ell_j^2 \times d$, where $d$ is the width of the UCS ($ \sim 10^5\,$cm), we estimate easily the length $\ell_j$ of the current sheet,
\begin{equation}\label{acc}
\ell_j=\sqrt{\frac{V_j}{d}}
\end{equation}
where the volume $V_j$ follows the probability distribution estimated from Fig.   \ref{volume}
\begin{equation}\label{vol}
P(V_j)=AV_j^{-a}, \;\;\; V_j^{min}< V_j< V_j^{max},
\end{equation}
where $A$ is the normalization constant, 
and $a=2.31$, $V_j^{min}=0.4\,(\mathrm{Mm})^3, V_j^{max}=130\,(\mathrm{Mm})^3$ (the extremes of the distribution - see Fig.\ref{volume}).
Combining Eqs. \ref{acc} and \ref{vol} we can estimate the UCSs'
random length $\ell_j$.

\paragraph{Electric field strength.$-$}\label{ef}
The electric field along the magnetic field, as we mentioned already, inside the UCS is
\begin{equation}\label{Ohm}
E=\eta \left\langle j\right\rangle
\end{equation}
where $\left\langle j\right\rangle$ is the current given by the probability distribution $P(\left\langle j\right\rangle)$ shown in
Fig.~\ref{current}. The resistivity is assumed close to zero when $\left\langle j\right\rangle<j_{\rm th} \approx 1580\,\mathrm{stA}\,\mathrm{cm}^{-2}$ and $\eta \approx \bar{\eta} \;\eta_S$ when $\left\langle j\right\rangle >j_{\rm th}$ (see Table \ref{tab2}), where $\eta_S $ is the Spitzer resistivity
\begin{equation}\label{spitzer}
\eta_S=\frac{m_e \nu_e}{n e^2}
\end{equation}
and $\bar{\eta}$ is a free parameter.
By including $\bar{\eta}$, we implicitly assume that due to the relatively strong current ($j>j_{\rm th}$) low-frequency electrostatic waves are excited and the particles interact with the waves much more efficiently than via Coulomb collisions, so the resistivity is enhanced by several orders of magnitude and is called `anomalous' \citep{Sagdeev, Papadopoulos, Galeev, Labelle, Ugai, Petkaki}. According to the literature stated above, $\bar{\eta}$ is proportional to $(j-j_{\rm th}).$ Since in our study, however, the driven current exhibits a steep gradient with a variation spanning less than one order of magnitude (see Fig. \ref{current}), we may assume a constant value for $\bar{\eta}$. The snapshot we have chosen represents a quiescent phase of the AR, 
and our aim is to investigate particle heating, so we will consider the resistivity to be only moderately increased above the Spitzer resistivity, choosing $\bar{\eta}$ around 75-100 (for the case where no collisions are taken into account). Note that these relatively low values are in contrast to the much higher values usually adopted \citep{archontis13} in the study of phenomena such as explosive events.
\paragraph{Equations of Motion.$-$}\label{trajectory}
We assume that the motion is one dimensional along the magnetic field lines and the velocity of the particles is non relativistic. The motion of the particle is divided into two parts:

(a) Free travel
along a distance $s_i$, suffering only collisional losses,
where we apply the simplified model of \cite{Lenard58} for
the Coulomb collisions of charged particles with a background plasma population
of temperature $T_b$,
\begin{equation}\label{loss1}
  \frac{ds}{dt}=u
\end{equation}
\begin{equation}\label{loss}
 \frac{du}{dt}=- \nu_{e} u + \left (\sqrt{2\nu_{e}k_B T_b/m_j} \right )\, W_t
\end{equation}
where $k_B$ is the Boltzmann constant, $m_j$ the particle (ion/electron) mass, and $W_t$
is an independent Gaussian random variable with
mean value zero and variance equal to the integration time-step $\Delta t$.
Equations (\ref{loss1}) and (\ref{loss}) are solved by directly using the analytical solution 
$s_a(\tau;s_0,u_0)$ and $u_a(\tau;u_0)$ (with $s_0$ and $u_0$ the values of the position and velocity at $\tau=0$) given in Gillespie (1996): for a prescribed free travel distance $s_i$, we first calculate the total free travel time $\tau_i$ by solving the nonlinear equation $s_a(\tau_i; s_0=0,u_0=u(t))=s_i$, and then determine the new velocity as $u(t+\tau_i)=u_a(\tau_i;u_0=u(t))$ in one step. This method allows the collision model
to be more realistic in that the collision frequency can be made
proportional to $1/u^3$, with the characteristic reduced collisionality
at high velocities.

(b) The particle is energized  by the UCS of length $\ell_j$ (Eq. \ref{acc})
\begin{equation}\label{space2}
  \frac{ds}{dt}=u
\end{equation}
\begin{equation}\label{cs}
\frac{du}{dt}=(e/m_j) \cos(\alpha) E
\end{equation}
with $\alpha$ the angle between the magnetic and the electric field, which is assumed to be random.

We release $10^5$ ions or electrons inside the AR with initial
velocity that obeys a Maxwellian
distribution with temperature $T\approx 10\ {\rm eV}$, and follow the evolution of their
velocity distribution in time, using the parameters 
determined earlier.
\begin{figure*}[!ht]
\centering
\subfloat[][]{\includegraphics[width=0.48\columnwidth,clip]{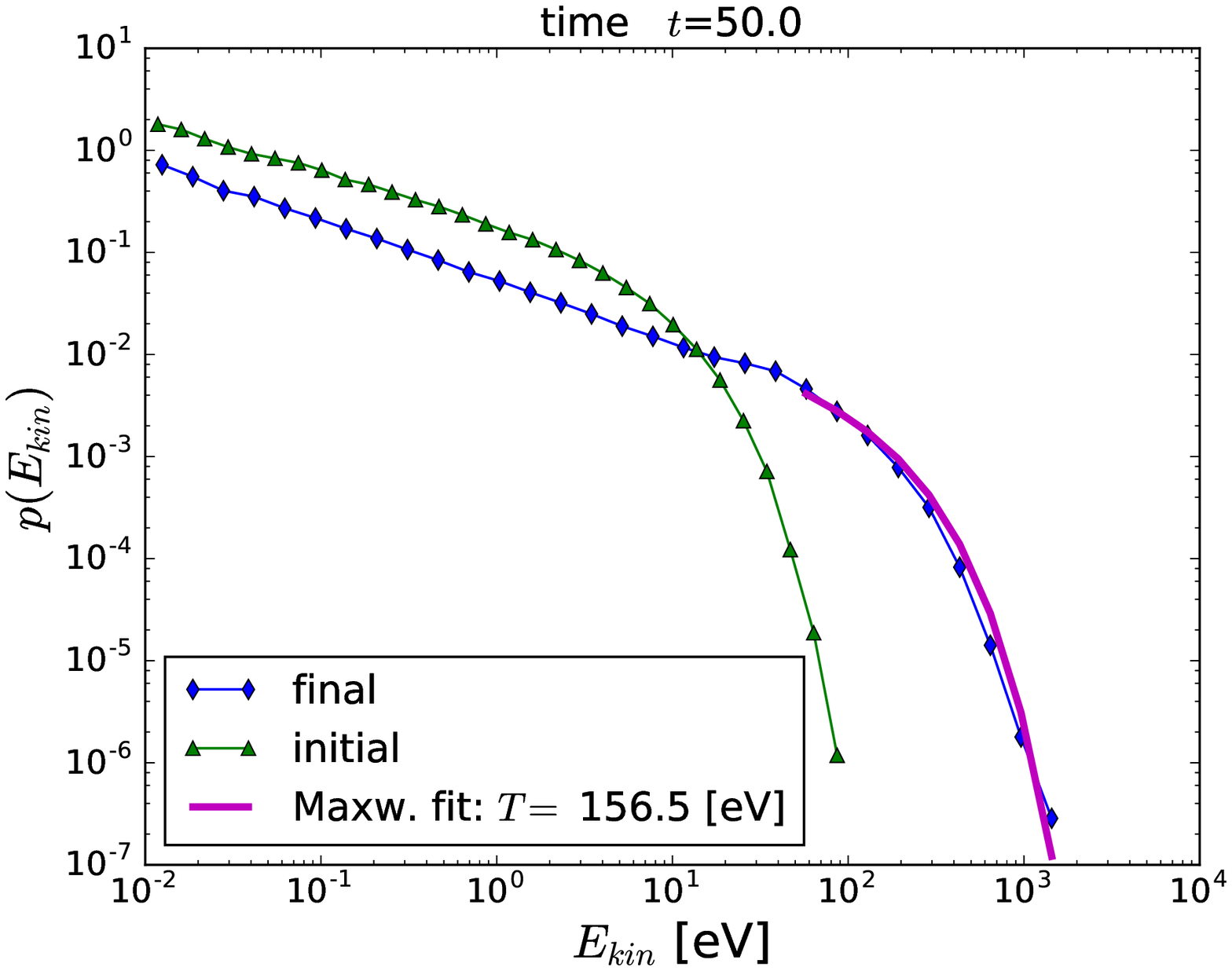}%
		\label{fig:9a}}
\subfloat[][]{\includegraphics[width=0.48\columnwidth,clip]{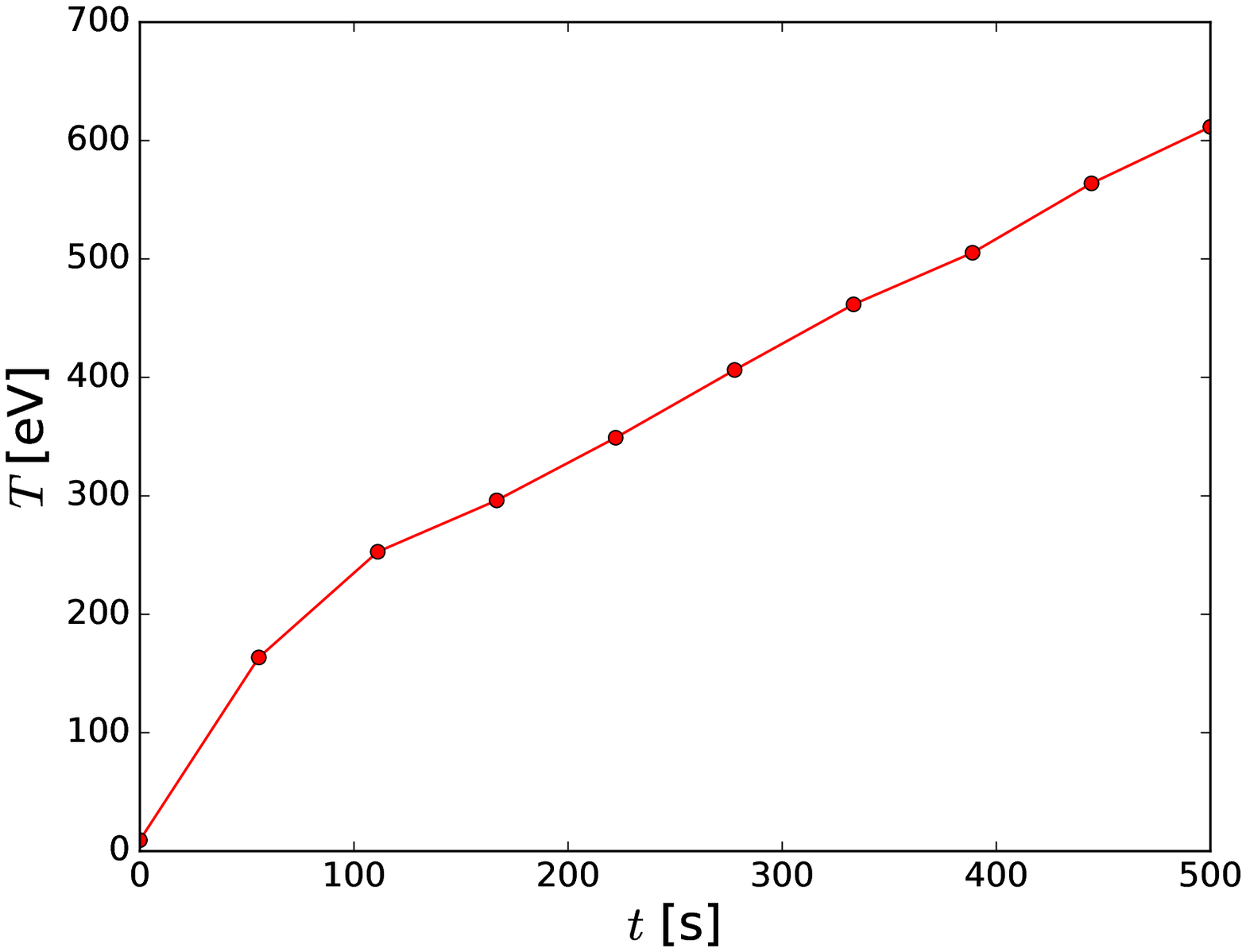}%
		\label{fig:9b}}\hfill\\
\subfloat[][]{\includegraphics[width=0.48\columnwidth,clip]{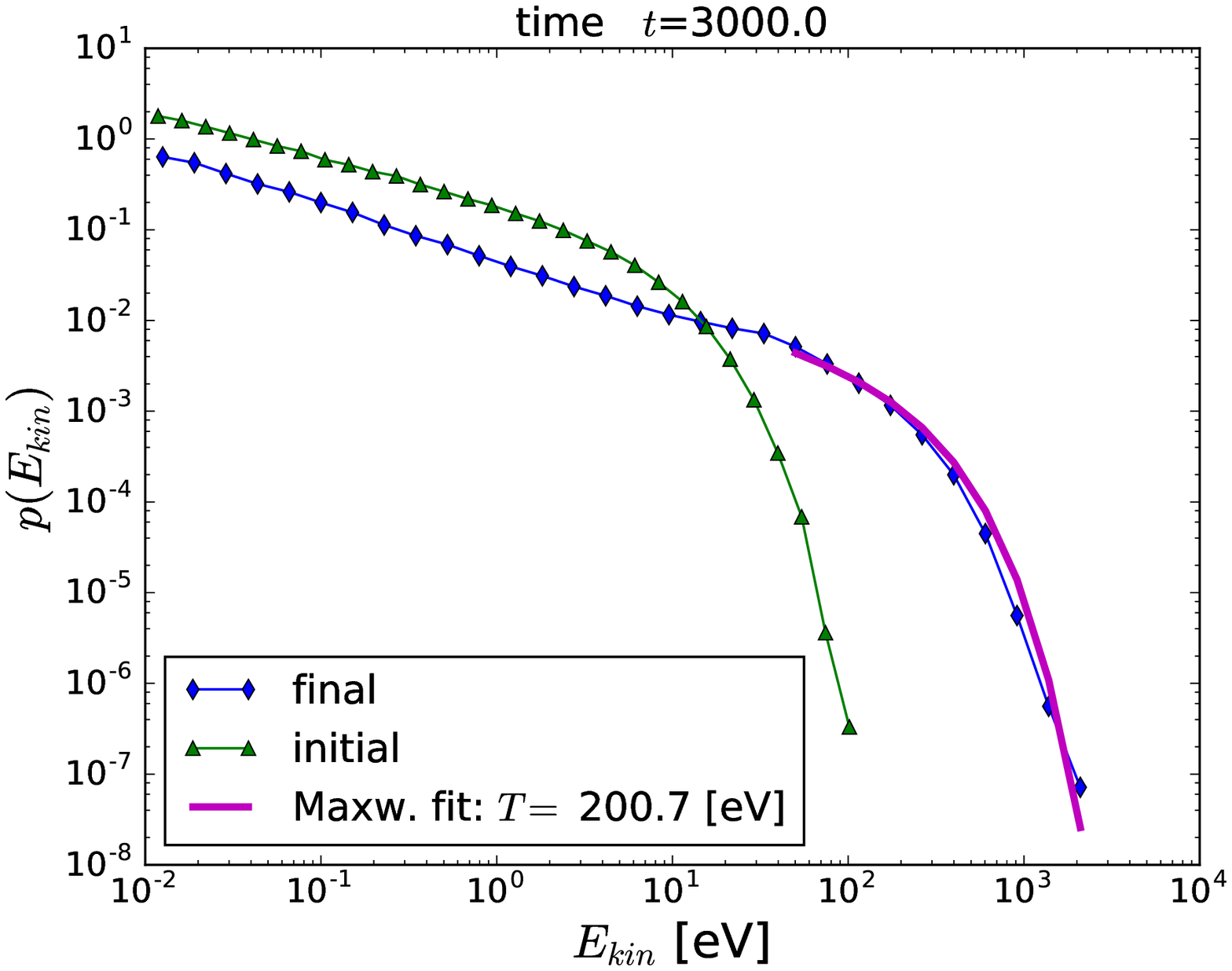}%
		\label{fig:9c}}
\subfloat[][]{\includegraphics[width=0.48\columnwidth,clip]{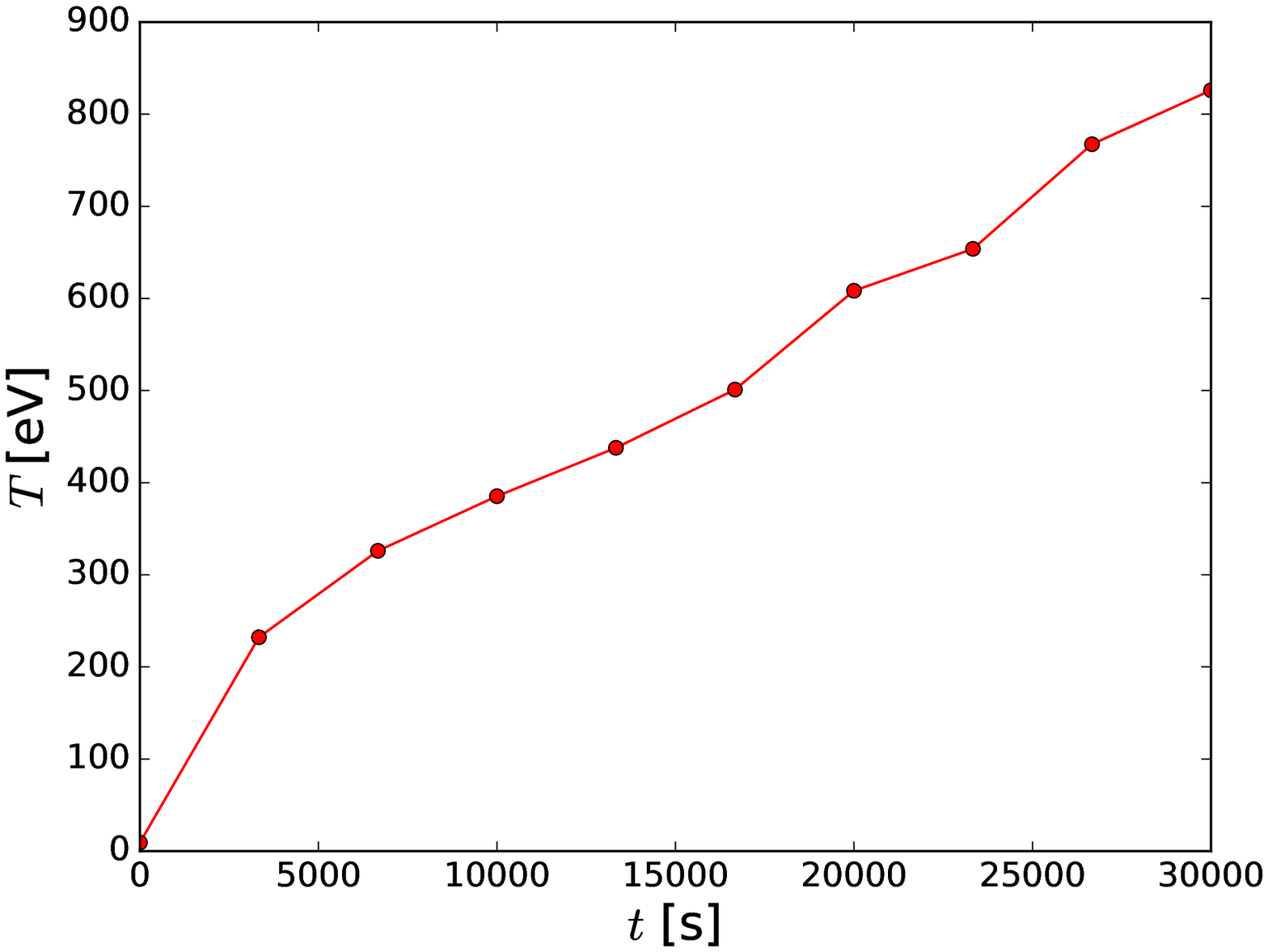}%
		\label{fig:9d}}
\caption{For $\eta= 75\eta_S$, we show: \protect\subref{fig:9a} Distribution of the kinetic energy of electrons 
after 50~s in the absence of collisions, 
together with a Maxwellian fit; \protect\subref{fig:9b} Electron temperature as a function of time in the absence of 
collisions, as estimated by Maxwellian fits; \protect\subref{fig:9c} Distribution of the kinetic energy of ions 
after 3000~s in the absence of collisions, 
together with a Maxwellian fit; \protect\subref{fig:9d} Ion temperature as a function of time in the absence of 
collisions, as estimated by Maxwellian fits.}
\label{ncollisions}
\end{figure*}

In Fig.\ \ref{ncollisions}a we show the
kinetic energy distribution of electrons at $50\,$s, for $\eta = 75 \eta_S$, 
in case of no collisions. The particles can escape in the vertical direction
($z$-) (the box size is $10^{10}$cm) but usually the number of escaping particles is small,
so they are not shown here.
All intermediate and final distributions are of clear Maxwellian shape, with particles
heated to a temperature of $150\,$eV.
As Fig.\ \ref{ncollisions}b shows, the electron temperature continuously
increases with time, there is no feedback or saturation mechanism since we
have not considered collisions.
Fig.\ \ref{ncollisions}c shows the kinetic energy distribution for
ions at $3000\,$s, for $\eta = 75 \eta_S$, without collisional effects. The
distribution again is of Maxwellian shape, and the
particles 
are heated to a temperature of $200\,$eV. Ions thus show a behaviour  similar
 to that of the electrons, but on a much slower time-scale, so they need
longer times for the heating mechanism to act effectively. More specifically,
the ions reach a temperature of $150\,$eV, which electrons have acquired
after $50\,$s, at a time roughly $2100\,$s.
This implies a scale factor for the
energization time that is close
to the square root of the proton-to-electron mass ratio.
Also, as for electrons, there is no saturation effect for ions, with temperature monotonically increasing.

In order for the electric field to be competitive against collisions,
the convective loss term in Eq.\ (\ref{loss}) should be smaller
than the electric force term in Eq.\ (\ref{cs}). Equating the two terms we find that, if the electric force is to dominate over collisonal losses, $\eta$ should be larger than $10^6 \eta_S$.
We thus consider here the low-resistivity regime, and
collisions must be considered important for the case of the quiescent snapshot we study.

\begin{figure*}[!ht]
\centering
\subfloat[][]{\includegraphics[width=0.48\columnwidth,clip]{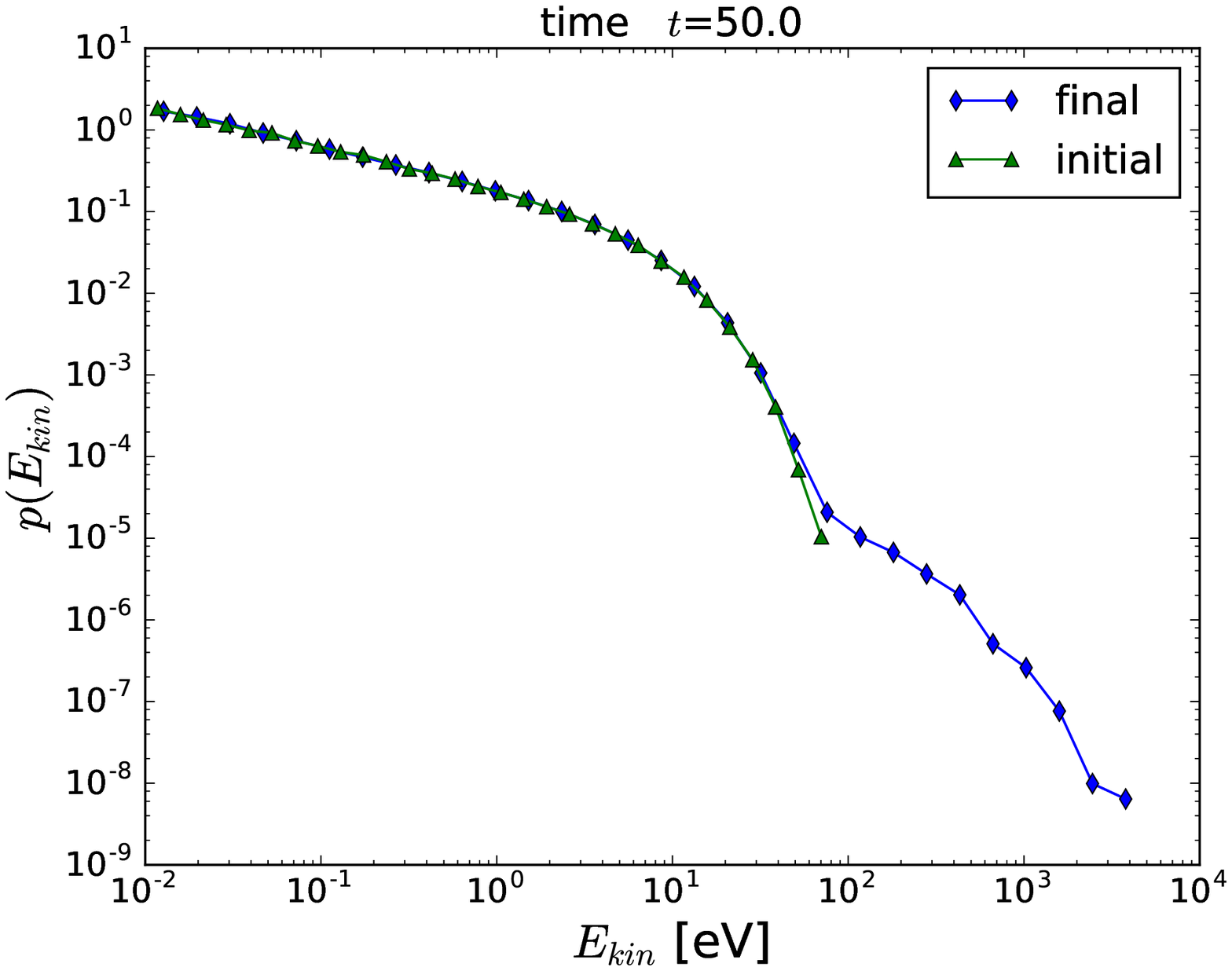}%
		\label{fig:10a}}
\subfloat[][]{\includegraphics[width=0.48\columnwidth,clip]{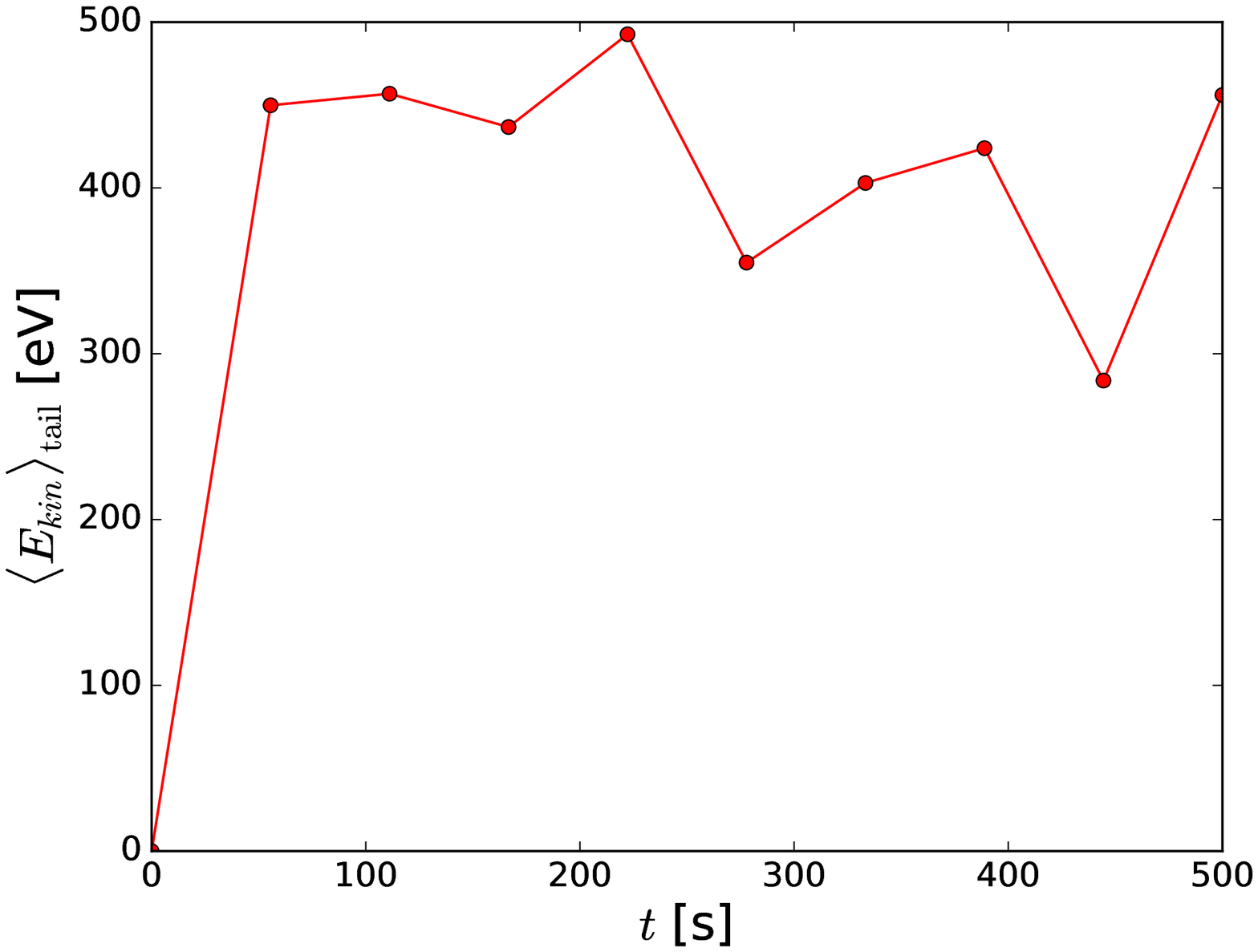}%
		\label{fig:10b}}\hfill\\
\subfloat[][]{\includegraphics[width=0.48\columnwidth,clip]{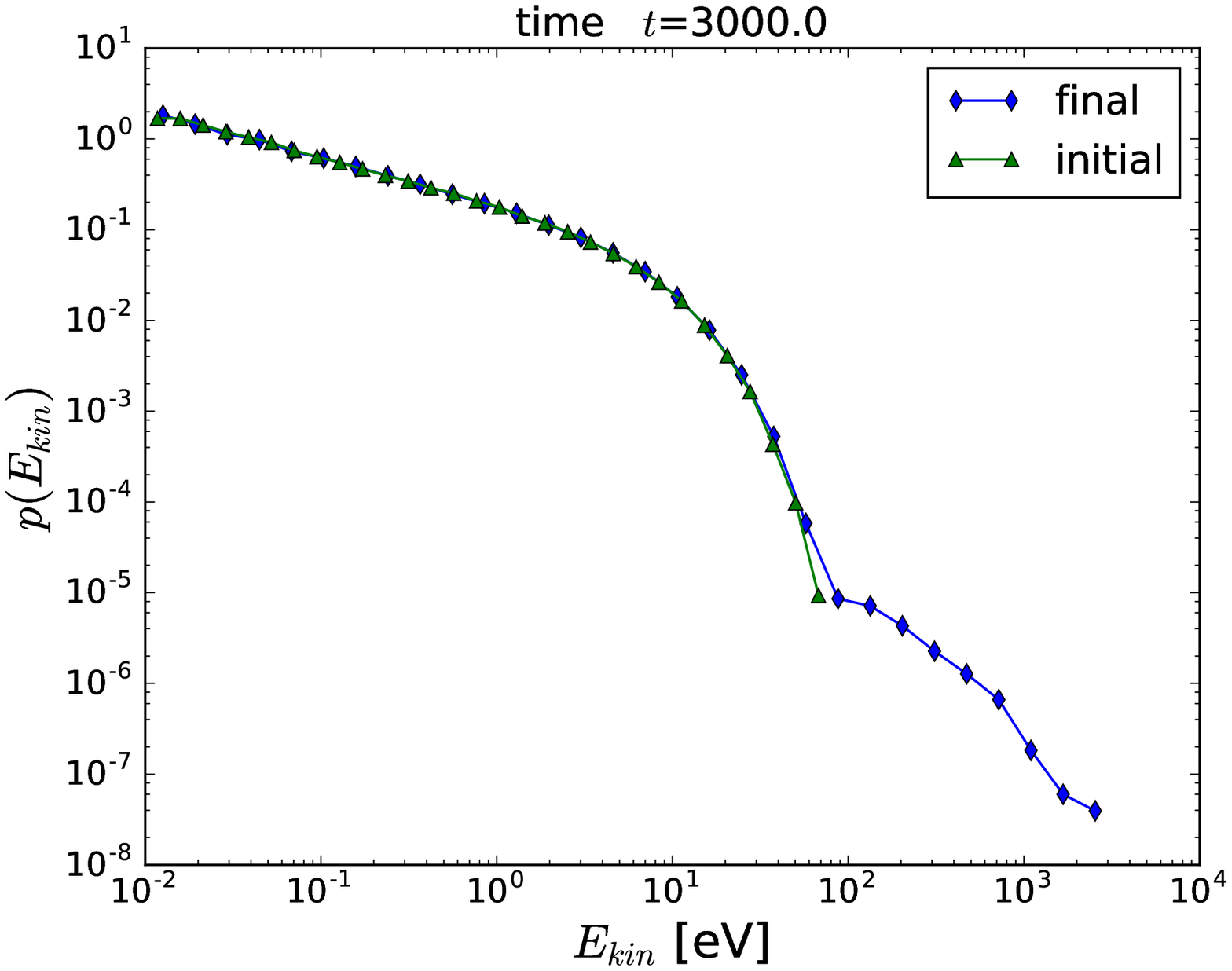}%
		\label{fig:10c}}
\subfloat[][]{\includegraphics[width=0.48\columnwidth,clip]{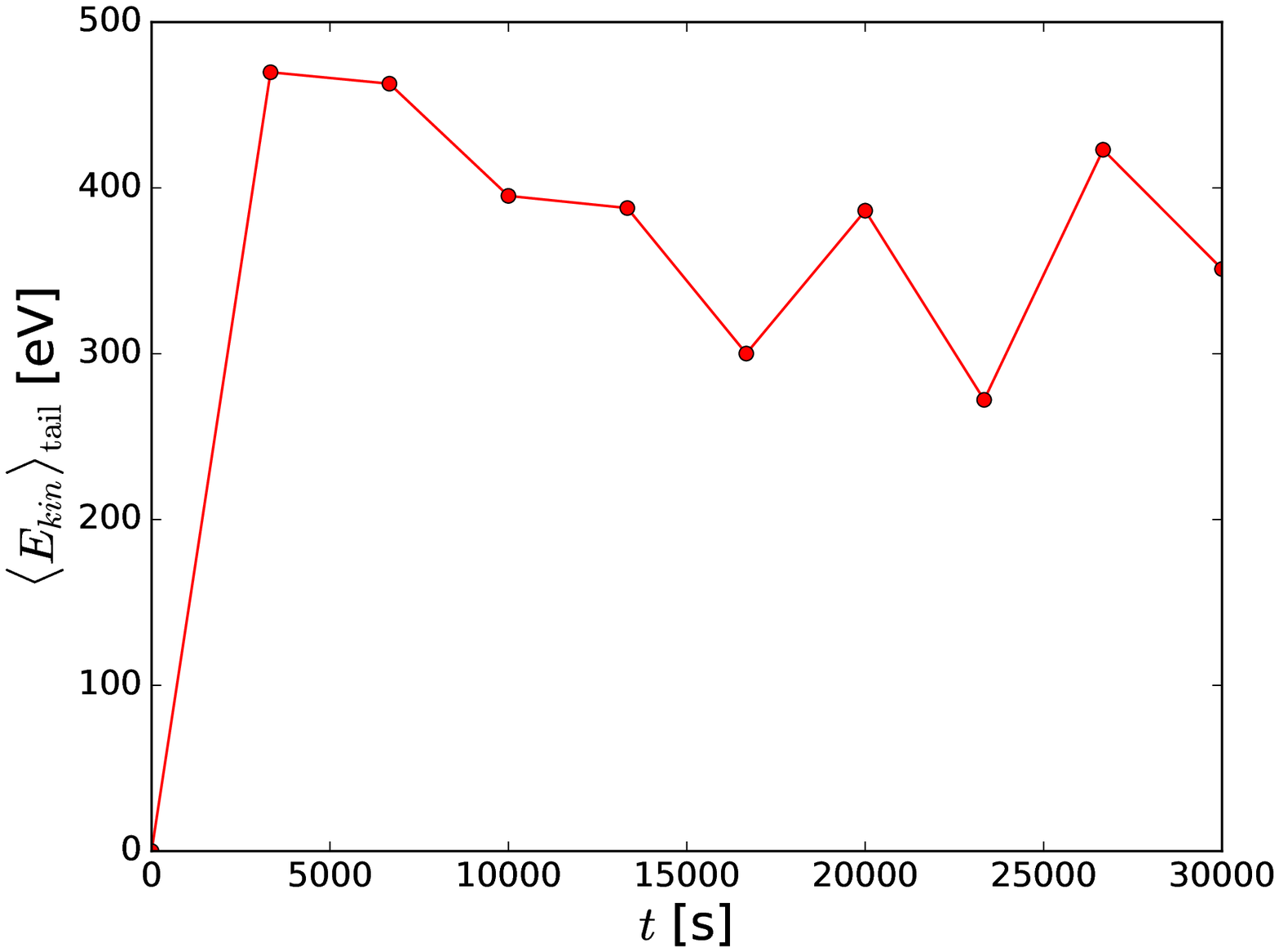}%
		\label{fig:10d}}
\caption{
Distribution of the kinetic energy with collisions included, for $\eta = 750 \eta_S$, 
\protect\subref{fig:10a} for the electrons at $t=50\,$s, and \protect\subref{fig:10b} the mean value of the kinetic energy of the electrons that form the power law tail (i.e. with energies larger than 100 eV) as a function of time, \protect\subref{fig:10c} for the ions at $t=3000\,$s, and \protect\subref{fig:10d} the mean value of the kinetic energy of the ions that form the power law tail (i.e. with energies larger than 100 eV) as a function of time.
}
\label{collisions}
\end{figure*}

We now apply the collisional losses (Eq.\ (\ref{loss})) during the free
travel times.
Fig.\ \ref{collisions}a shows the kinetic energy distribution
of electrons at $50\,$s for $\eta=750\eta_S$, with collisions included,
and Fig.\ \ref{collisions}c shows the respective distribution
for ions at $3000\,$s and for $\eta=750\eta_S$.
The bulk of the  particles
lose all the energy they gain to the
background, and a small fraction of the particles forms a power-law tail, though not much extended, for both electrons and ions. 
We also find that a higher value of $\eta$ (we here used already the 10 times
higher value of $\eta = 750 \eta_S$ compared to the value of $\eta = 75\eta_S$ used otherwise in this article) does not alter the
behaviour of the bulk population, it leading only to a more extended tail.
So we do not find any heating of the test particle population when collisions
are included, which can though readily be explained by the fact that
even the minimum free travel distance $s_{\rm min}=10^8\,$cm
is much larger than the collisional mean free path ($10^4-10^5\,$cm),
so the majority of the particles undergo a very
large number of collisions. That the test-particles are not heated implies that there is a transfer of energy to the background population, i.e. the bulk population would be heated, which is  though not taken into account in our modelling approach. The tail of both particle species
can be explained by the fact that the collision frequency depends on the third power of the
inverse instantaneous velocity, and with that fast particles are much less
affected by collisions. The total energy carried by the tails 
remains roughly stable over time (see Fig.\ \ref{collisions}b, d).

\begin{figure*}[!ht]
\centering
\subfloat[][]{\includegraphics[width=0.48\columnwidth,clip]{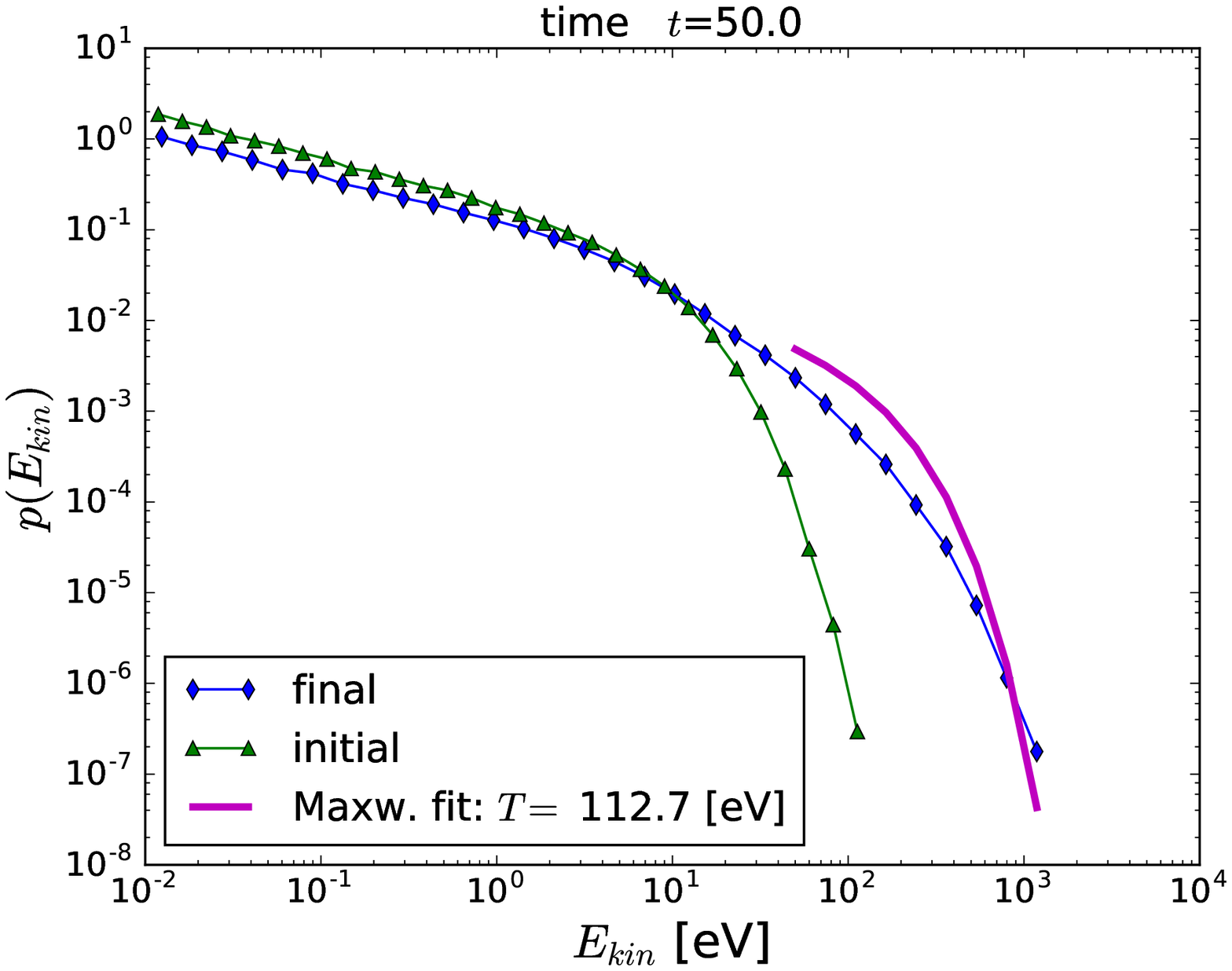}%
		\label{fig:11a}}
\subfloat[][]{\includegraphics[width=0.48\columnwidth,clip]{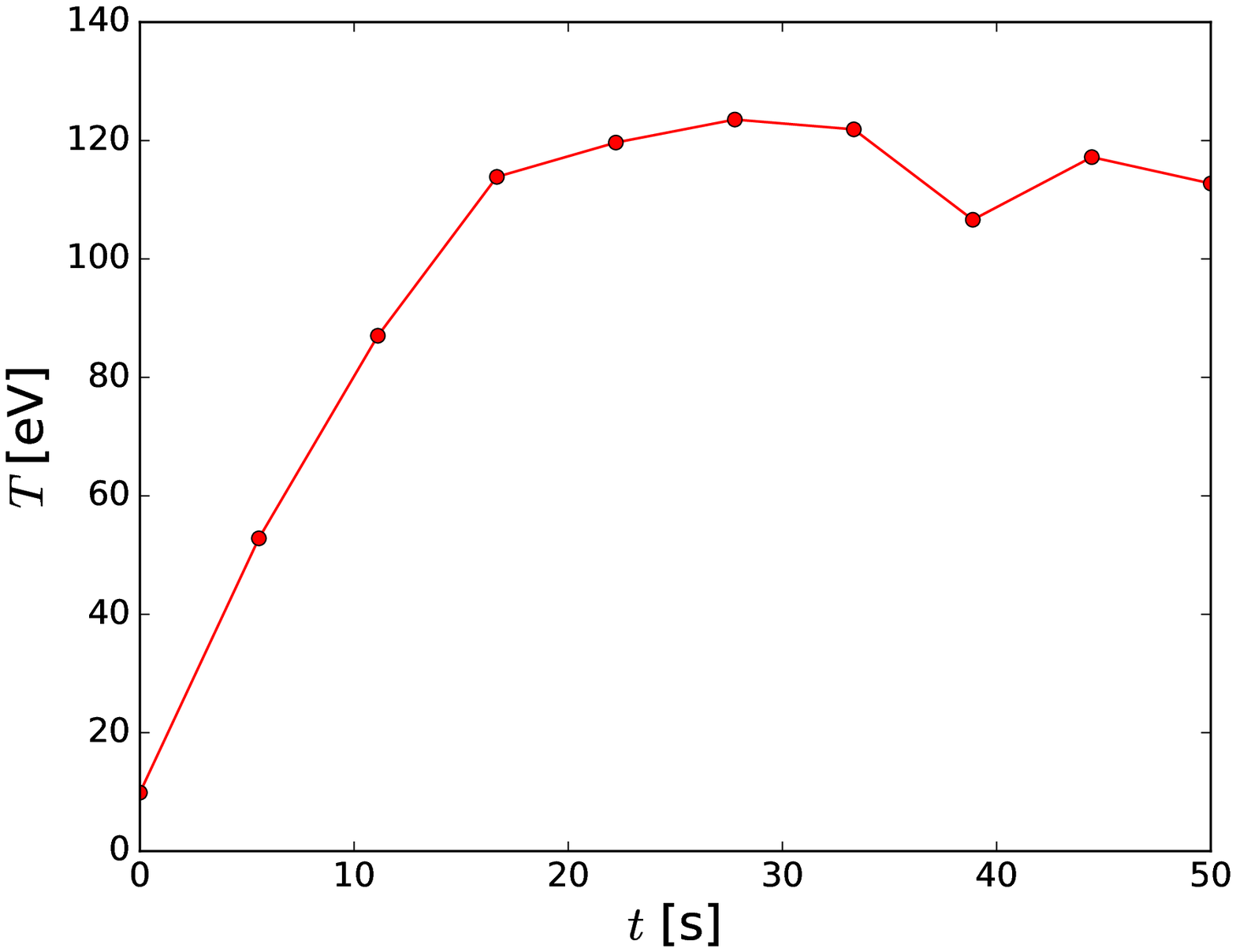}%
		\label{fig:11b}}\hfill\\
\subfloat[][]{\includegraphics[width=0.48\columnwidth,clip]{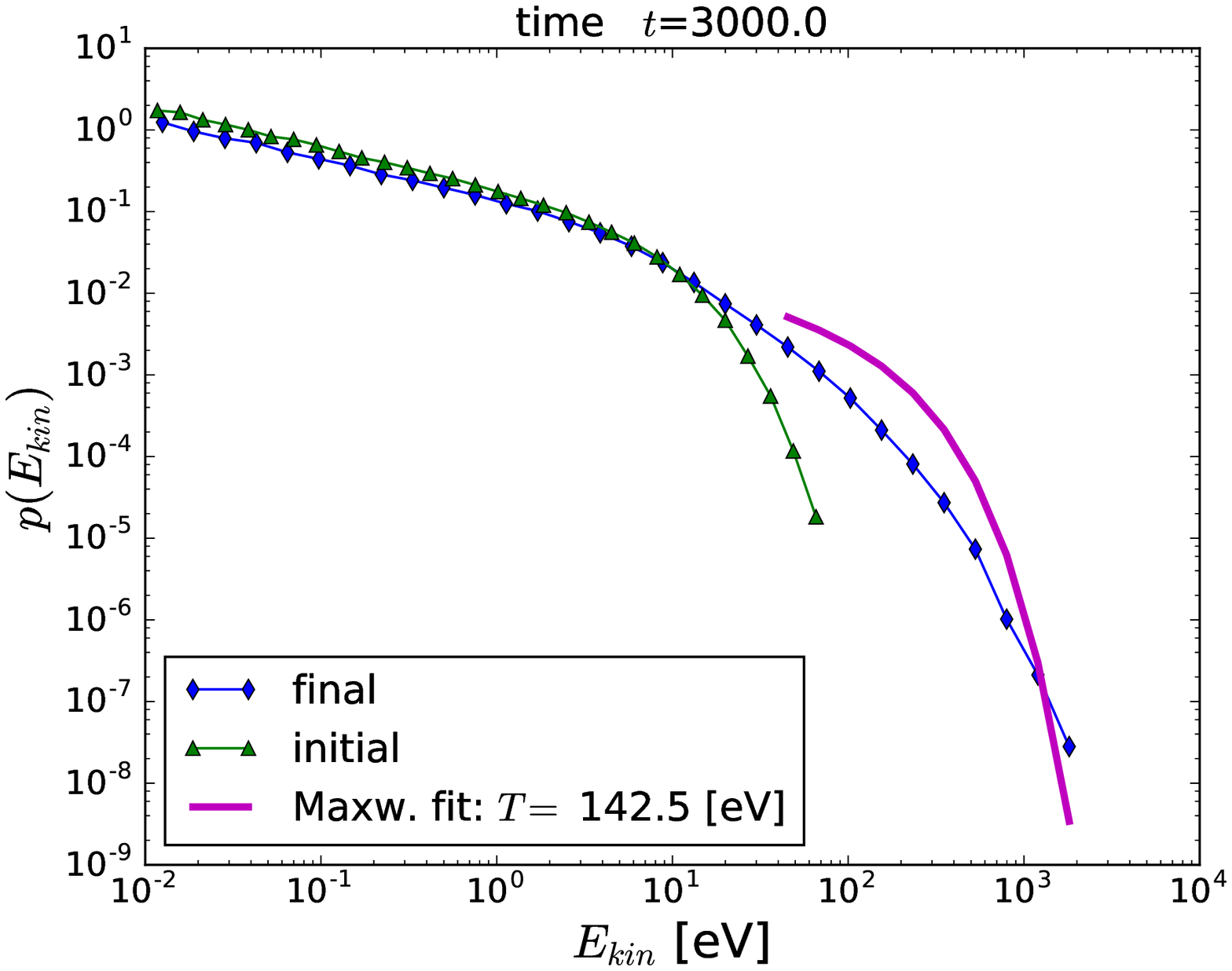}%
		\label{fig:11c}}
\subfloat[][]{\includegraphics[width=0.48\columnwidth,clip]{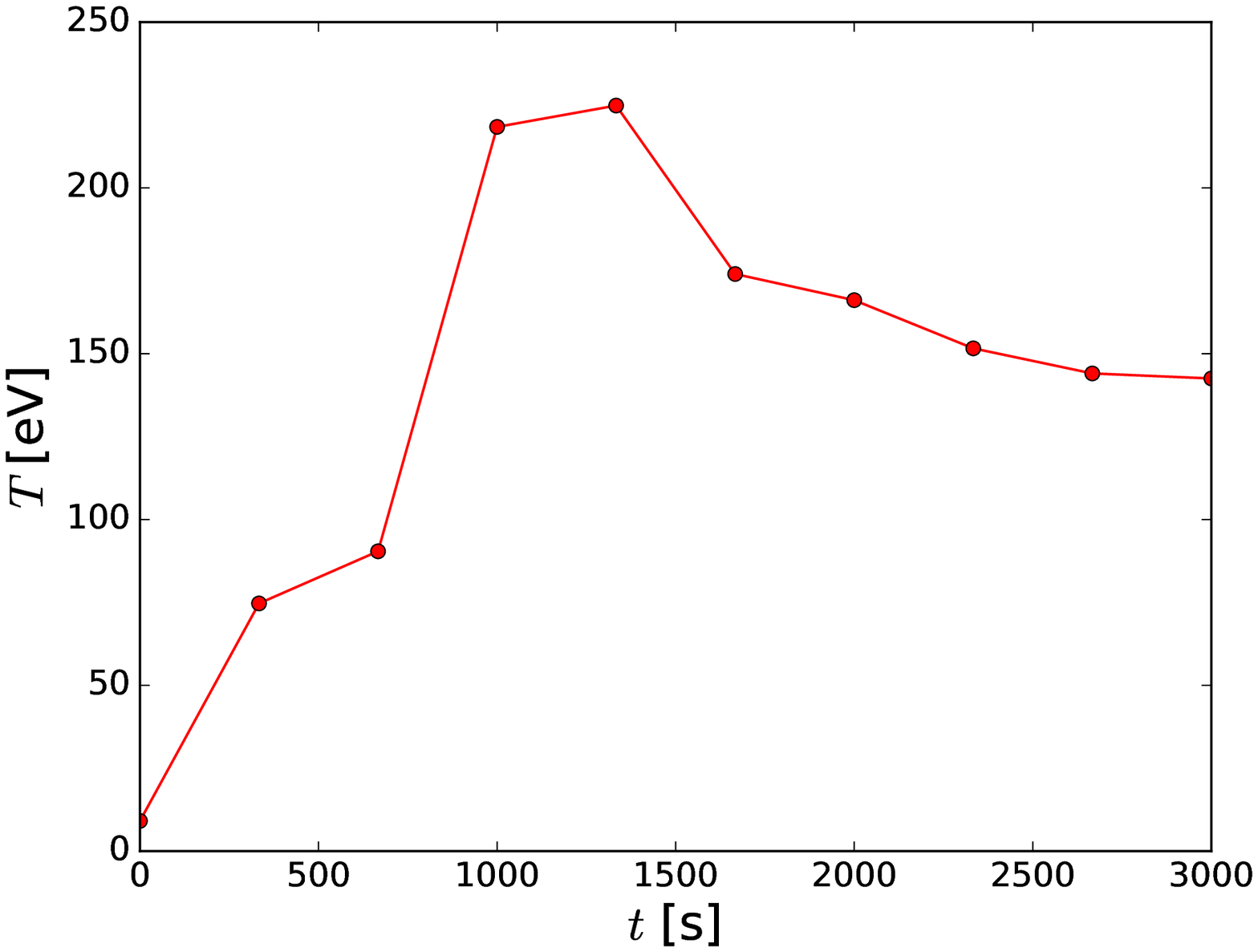}%
		\label{fig:11d}}
\caption{Same as Fig. \ref{ncollisions}, but including the effect of collisions, and with a modified (smaller) value of $s_{\mathrm{min}}$.}
\label{mcollisions}
\end{figure*}

The value of $s_{min}$ was so far chosen to be close to the grid-resolution
of the extrapolated field, and with that it is more a technical than
a physical limit. In the following, we make one single change to
the parameters as obtained from the data analysis, by setting $s_{\rm min}=10^4\,$cm,
i.e.\ we let it be of the order of the mean free path. The aim of this change is to explore whether, depending on the scales of the current fragmentation, 
a kinetic description may become appropriate.
The resulting kinetic energy distribution of electrons is shown in
Fig.\ \ref{mcollisions}a, and obviously it is close to, but not exactly of, Maxwellian shape, with temperature roughly $110\,$eV for $\eta = 75 \eta_S$.
As Fig.\ \ref{mcollisions}b shows, a saturation occurs at this
enhanced temperature. We also find that 
for the larger resistivity $\eta = 150\eta_S$ there is again heating with saturation at a temperature twice as much as for $\eta=75\eta_S$ 
(not shown here).

In Fig.\ \ref{mcollisions}c, the kinetic energy distribution of ions
is shown
at $3000\,$s for $\eta=75\eta_S$, again close to, but not perfectly, of Maxwellian shape, and from Fig.\ \ref{mcollisions}d it follows that
there is again saturation of the heating process at roughly $140\,$eV,
nearly the same temperature as the one of the electrons.

\section{Summary}

This work discusses an observationally driven study of magnetic configuration, particle acceleration and plasma heating pertaining to NOAA AR 11158. The analysis consists of the following steps: 

Using nonlinear force-free extrapolation techniques we reconstructed the magnetic field of the active-region corona using several choices for the spatial-resolution. We searched for magnetic discontinuities and UCS built into the reconstructed magnetic field, at the same time performing a validation of the extrapolated fields aiming to discard structures with poor field solenoidality. 

The next step was the statistical analysis of UCS and the search for the characteristics of the spatial clustering and strength of their electric currents. We found that the clustering shows stable power-law behaviour for (a) the electric current density distribution above a certain threshold, (b) the magnetic energy distribution of the UCS, and (c) their volume distribution. In addition, UCSs were found to be fractal, with a well-defined fractal dimension, $\sim 1.8$. Notice that the power-law distributions for UCS free energies and volumes align with those of previous estimations using linear force-free results \citep{Vlahos04}.

The statistical results, most notably the power-law indices, were found to be fairly insensitive to the spatial resolution. This conclusion was reached by testing much higher-resolution data from \textit{Hinode}.

Based on the above statistical characteristics and using a small enhancement for the resistivity above the Spitzer resistivity, we reconstructed the fractal distribution of the electric field inside the active region. 

We followed $10^5$ plasma particles inside the fractal electric fields and monitored the temporal evolution of their kinetic energy distribution for both electrons and ions.

We found that, due to collisions, both electrons and ions are energized only at the tail of their final distribution, that is not necessarily identical to their initial Maxwellian. This tail typically obeys an energetically stable power-law whose dynamical range depends on the value of the resistivity.

We conclude that the bulk of the plasma is heated either directly or collisionally:\\
(a) If we assume that the smallest distances between UCS are given by the resolution of the magnetogram and its extrapolation, then collisions dominate such that the heating can only be modelled by using a standard fluid transport approach.\\ 
(b) If the smallest distances between the UCS are of the order of the collisional mean free path, then heating is manifested on the kinetic level adopted in this article. Therefore, the appropriate approach to the question of coronal heating, namely fluid or kinetic level, depends on the scales to which the current fragmentation extends.

Assuming that the same power-law distribution of UCS distances extends below the spatial resolution of the extrapolated magnetic fields we found that, close to the collisional mean free path, both electrons and ions reach temperatures in excess of 100-150 eV in a few tens of seconds for electrons and a few thousands of seconds for ions. Therefore, our kinetic analysis shows that if nanoflares with these properties exist in the active-region corona, they should be able to easily heat the plasma to millions of K. This may conceivably hold for the quiet-Sun corona, as well, even with fewer or weaker UCS, given the lower energy-loss demands away from active regions.

Making the conjecture that the statistical properties of the UCS in quiescent ARs generally remain unchanged from what we found here, the plasma in the solar corona can be heated either through the interaction of the plasma particles with UCS (following the kinetic approach, see Fig.~\ref{mcollisions}), or through a combination of collisional local heating (current dissipation, $\eta j^2$) and heat transport \citep{Klimchuk2006,Klimchuk2015}. In the latter case, the statistical properties of the UCS can be used in order to distribute the dissipation regions inside the active region. Our model is a first attempt to investigate the role of kinetic processes in coronal heating. To fully address the problem, one would need to include the full behavior of the plasma, most notably the role of cooling processes. Put simply, the cooling rate of the plasma should balance the rate of energy (heat) deposition at quasi-steady temperatures compatible to those of the ambient solar corona. The inclusion of plasma, either via a fully kinetic code or via a hybrid one, comprising a hydrodynamical treatment of the low-energy particles and a kinetic treatment for the high-energy ones, is a viable extension to this work.

On the statistical aspect, if UCS are to be associated with coronal heating, then the sub-resolution scales and the evolution (fragmentation and coalesence) of the large scale UCS, analyzed in this study holds the secret. Our power-law distribution of the UCS energetics (Fig.~\ref{energy}) will dynamicaly evolve and the satistics of the sporadic heating may be very different, especially on small scales. Aimining to as high spatial resolution as feasible will lead us to the kinetic analysis of the coronal heating problem and this will be another meanigful extention of this study.
	
Concluding, we have shown that the obesrvationally-driven reconstruction of the coronal magnetic field is a meaningful approach for studies of active-region coronal heating by fragmented currents. Our analysis has two drawbacks, namely (1) the low spatial resolution of the reconstructed magnetic field ($\approx 700 \; \rm{km}$), which is much coarser than the collisional mean free path ($\approx 1\; \rm{km}$) and (2) the cursory use of the anomalous resistivity, whose role should be investigated comprehensively for weak currents. We plan to expand on these two research avenues and combine our analysis with currently available one-dimensional techniques such as hydrodynamic-fluid codes for coronal heating \citep[see][]{Klimchuk2006, Klimchuk2015, Cargill15} and UCS in more realistic, MHD rather than force-free, magnetic configurations \citep[e.g.][]{Archontis}.

\begin{acknowledgements} 
We thank Drs A. Anastasiadis and S. Patsourakos for reading the article and making valuable suggestions. The authors acknowledge support by European Union (European Social Fund-ESF) and Greek national funds through the Operational Program Education and Lifelong Learning of the National Strategic Reference Framework (NSRF)-Research Funding Program: Thales: Investing in knowledge society through the European Social Fund.
\end{acknowledgements}

\end{document}